\documentclass[12pt]{iopart}
\usepackage{iopams}
\usepackage{color}
\usepackage{bm}
\usepackage{graphicx}
\usepackage{tikz}
\begin{document}

\title{Power-Duality in Path Integral Formulation of Quantum Mechanics}

\author{Akira Inomata}
\address{%
Department of Physics, State University of New York at Albany,\\
Albany, NY 12222, USA
}%
\ead{ainomata@albany.edu}
\author{Georg Junker}
\address{Institut für Theoretische Physik I, Friedrich-Alexander-Universität Erlangen-Nürnberg,
Staudtstr.\ 7, 91058 Erlangen, Germany \\[2mm] and\\[2mm]
European Organization for Astronomical Research in the Southern Hemisphere,\\
Karl-Schwarzschild-Stra\ss e 2, D-85748 Garching, Germany
}%
\ead{georg.junker@fau.de;gjunker@eso.org}

\begin{abstract}
Power duality in Feynman's path integral formulation of quantum mechanics is investigated. The power duality transformation consists of a change in coordinate and time variables, an exchange of energy and coupling, and a classical angular momentum replacement. Two physical systems connected by the transformation form a power-dual pair.
The propagator (Feynman's kernel) expressed by Feynman's path integral cannot be form-invariant under the transformation, whereas the promotor constructed by modifying Feynman's path integral is found form-invariant insofar as the angular momentum is classical. Upon angular quantization, the power duality breaks down. To save the notion of power duality, the idea of quasi power duality is proposed, which constitutes of an ad hoc angular momentum replacement.  The power-dual invariant promotor leads to the quasi-dual invariant Green function. A formula is proposed, which determines the Green function for one of a dual pair by knowing the Green function for the other. As examples, the Coulomb-Hooke dual pair and a family of two-term confinement potentials for a zero-energy state are discussed.
\end{abstract}

\section{Introduction}\label{Sec1}

In an earlier article \cite{IJ2021}, we have studied the duality between two power force laws (power duality in short) in classical, semiclassical and quantum mechanics.  In the present paper, we wish to investigate the power duality in Feynman's path integral formulation of quantum mechanics \cite{Feynman1948,Feyn,Schu}.

In recent years, numerous exoplanets have been discovered \cite{ExoPlanets}. It is a generally accepted view that Newton's law of gravitation holds in extrasolar systems. Orbital mechanics of exoplanets, as is in the case of  solar planets and satellites, mainly deals with the Kepler problem with perturbation. The common procedure for studying perturbations to the Kepler orbit is the so-called regularization, introduced by Levi-Civita \cite{LC1906} and generalized by Kustaanheimo and Stiefel \cite{KS1965}. The regularization is a way to transform the singular Kepler motion to the non-singular Hooke motion. There are many ways to convert Newton's law to Hooke's law via various transformations of real numbers, complex numbers, spinors, quaternions, hypercomplex numbers and so on. See references \cite{Arnold,Chandrasekhar,Needham1993,Needham1997,Grant1994,Kasner,Hall,Grandati,Grandati2010,Vivarelli,Vrbik}.
These efforts on regularization is indeed all based on recognition of the dual relation between Newton's law and Hooke's law.
In the above,  by Newton's law we mean the inverse-square force law and by Hooke's law the linear force law,
despite some arguments among historians that Hooke knew the inverse-square force for the Kepler motion prior to Newton's Principia (see, e.g., \cite{Nauenberg,Chin}).

Duality between the inverse-square force law and the linear force law seems to have been known to Newton and Hooke.
According to Chandrasekhar's reading \cite{Chandrasekhar} out of propositions and corollaries in Principia, Newton had even established duality between the centripetal forces of the form $r^{a}$ and $r^{b}$ for various pairs $(a, b)$. The power-law duality for arbitrary power forces in classical mechanics was analyzed by Kasner \cite{Kasner}, Arnol'd \cite{Arnold}, and others \cite{Hall,Grandati,Grandati2010}.  In our previous work \cite{IJ2021} we expanded the domain of the dual pairs.

In quantum mechanics, the classical Newton-Hooke duality corresponds to the dual relation between the Coulomb system and the harmonic oscillator. Schr\"odinger \cite{Schroe1,Schroe2} initially solved his wave equation for the hydrogen atom and the harmonic oscillator separately, and later found a transformation connecting the two systems \cite{Schroedinger}. Schr\"odinger's equation for arbitrary power potentials was extensively studied by Johnson \cite{Johnson} and others \cite{Gazeau,GrantRosner}. Application to the confinement problem was discussed by Quigg and Rosner \cite{Quigg}, Gazeau \cite{Gazeau} and Steiner \cite{Steiner1986}.  Supersymmetric aspects of the Coulomb-Hooke dual relation in arbitrary dimensions are also discussed by Kosteleleck\'y et al.\ \cite{Kostelecky}, see also \cite{Sukumar}. It is interesting to mention that such power-laws appear to play useful roles in the Anderson localization of low-dimensional disordered systems \cite{Rodriguez,Syzranov}.

Feynman's path integral approach to quantum mechanics is successful in solving the quadratic potential problems, but fails to produce the solution for the hydrogen atom which once symbolized the success of Schr\"odinger's wave equation. However, if the action of the form of Hamilton's principal function in Feynman's path integral is replaced by that of Hamiltoin's characteristic function, then one can get a help from the Coulomb-Hooke duality to solve the modified path integral for the hydrogen atom \cite{DuruKleinert,HoIno,Inomata1984,Steiner1984}. Under the influence of the duality idea,  the connection between the path integral approach and the dynamical group approach was also clarified \cite{IKG1992}. Such studies motivate us to investigate the power duality in path integration by asking
whether the power duality is a valid symmetry or a broken symmetry; what quantities remain invariant if valid; how breaking occurs if broken; what benefits can be gained altogether.  For a  more detailed review of the historical backgrounds of the power duality see \cite{IJ2021}.

Power duality is a relation between two objects in motion holding fixed values of energy $E$ and angular momentum $L$ under the influence of the central power force law. The central potential due to the power force law is in general of the form,
\begin{equation}\label{Potential}
V_a(r)=\lambda_{a}r^{a} + \lambda_{a'}r^{a'} + \lambda_{a''}r^{a''} + \cdots + \lambda_{a^{(M)}}r^{a^{(M)}}
\end{equation}
where $\lambda_{a}$ is the coupling constant associate with $r^{a}=|\mathbf{r}|^{a}$, the $a$-th power of the magnitude of a position vector $\mathbf{r}\in\mathbb{R}^D$, and so on.
We treat the first term as the primary potential and the remaining terms as the secondary potential.
In the present work, we consider mainly the two-term potential $(M=2)$ with non-vanishing coupling constants $\lambda_{a}$ and $\lambda_{a'}$.
There is a set of reversible operations, called \emph{the power-duality transformation} and denoted by $\Delta$, that takes a system with a power potential to another system with a different power potential. The power-duality transformation of the primary potential induces the changes in the secondary potential. Suppose system $A$ and system $B$ are connected by $\Delta$, with
\begin{equation}\label{Potential2}
V_b(\rho)=\lambda_{b}\rho^{b} + \lambda_{b'}\rho^{b'} + \lambda_{a''}\rho^{b''} + \cdots + \lambda_{b^{(M)}}\rho^{b^{(M)}}.
\end{equation}
Then we say, the two systems are \emph{power-dual} to each other. The power-duality transformation $\Delta(\mathfrak{R}, \rmd\mathfrak{T}, \mathfrak{L}, \mathfrak{E}, \mathfrak{S})$ consists of the following reversible operations,
\begin{itemize}
  \item Power transformation: \, $r^{a} \rightarrow \rho^{b}$;
  that is,
  \begin{equation}\label{powtrans}
    \mathfrak{R}: \, \quad \, r=f(\rho) =C\rho^{\eta},
  \end{equation}
  where $\eta = - b/a$ with $(a+2)(b+2)=4$, \\
  and $C$ is a real constant having a dimension of  $\rho^{1- \eta }$.

  \item Non-integrable time transformation: \, $\rmd t \rightarrow \rmd s$;  or more specifically,
  \begin{equation}\label{dT}
    \rmd \mathfrak{T}: \, \quad \, \rmd t=C^{2}\eta^{2} \rho^{2\eta -2} \rmd s.
  \end{equation}

  \item Angular momentum transformation: \, $L_{a} \rightarrow L_{b}$; \, that is,
  \begin{equation}\label{angtrans}
    \mathfrak{L}: \, \quad \, L_b = \left|\eta \right| L_a.
  \end{equation}

  \item  Energy-coupling swapping: \, $E_{a} \rightarrow \lambda_{b}, \, \lambda_{a} \rightarrow E_{b}$; \, that is,
  \begin{equation}\label{rename}
    \mathfrak{E}: \, \quad \, E_{b} = - \eta^{2}C^{a+2} \lambda_{a}, \, \quad \, \lambda_{b} = - \eta^{2}C^{2} E_{a}\,.
  \end{equation}

  \item Change in the secondary potential: \, $\lambda_{a^{(i)}} \rightarrow \lambda_{b^{(i)}}$, \,
  \begin{equation}\label{second}
    \mathfrak{S}: \, \quad  \lambda_{b^{(i)}}= \left(\frac{2}{a+2}\right)^{2} C^{a^{(i)}+2}\lambda_{a^{(i)}}, \, \quad \, b^{(i)}=\frac{2(a^{(i)}-a)}{a+2}\,,
  \end{equation}
where $i=1,2, 3,..., M$.
\end{itemize}
Being more explicit \cite{IJ2021}, let
\begin{equation}\label{Wa}
  W^{(a)} = \int \rmd t \left[ \frac{m}{2}\left(\frac{\rmd r}{\rmd t}\right)^2 - \frac{L_a^2}{2mr^2} -V_a(r) +E_a \right]
\end{equation}
be Hamilton’s radial characteristic function associated with central potential (\ref{Potential}) for given angular momentum $L_a$ and energy $E_a$. Then above power-duality transformation $\Delta$ maps $W_a \to W_b$, where
\begin{equation}\label{Wb}
  W^{(b)} = \int \rmd s \left[ \frac{m}{2}\left(\frac{\rmd \rho}{\rmd s}\right)^2 - \frac{L_b^2}{2m\rho^2} -V_b(\rho) +E_b \right]
\end{equation}
is Hamilton’s radial characteristic function associated with central potential (\ref{Potential2}).
A brief historical review on power duality is given in reference \cite{IJ2021}.

Suppose quantities $Q_{A}(a,a')$ and $Q_{B}(b,b')$ belong to system $A$ and system $B$, respectively.
Suppose $Q_{A}(a,a')$ is taken to $Q_{B}(b,b')$ by $\Delta$, that is, $Q_{A}(a,a')=Q_{B}(b,b')$ via $\Delta$. Then we also say that $Q_{A}(a,a')$ and $Q_{B}(b,b')$ are power-dual to each other. Let $X(a,b)$ be an operation which exchanges the parameters $(a,a')$ and $(b,b')$. If $Q_{B}(b,b')$ returns to $Q_{A}(a,a')$ under $X(a,b)$, that is, if $X(a,b)Q_{B}(b,b')=Q_{B}(a,a')=Q_{A}(a,a')$, then we say that $Q_{A}(a,a')$ is \emph{dual-form invariant} under $\Delta$ and that $Q_{A}(a,a')$ and $Q_{B}(b,b')$ are \emph{symmetric} with respect to $\Delta$.

In classical mechanics, as was observed in the previous work \cite{IJ2021}, the power-duality transformation $\Delta$ takes the action function of a power potential system to the action function of another power potential system. Since the path integral is based on the classical action, it is reasonable to expect that the power duality would be compatible with the path integral. However, we have to recognize that path integration is a non-trivial calculation which has been explicitly carried out only for a limited class of potential systems.
Standard books on path integration \cite{Feyn,Schu} mostly deal with quadratic systems on rectangular coordinates. For the power duality argument, it is necessary to prepare the polar coordinate formulation of the path integral. Furthermore, we have to modify Feynman's path integral by taking the following important points into account. The action function treated in classical mechanics was Hamilton's characteristic function, whereas the action appearing in Feynman's path integral is of the form of Hamilton's principal function. Therefore, we have to modify the path integral so as to be based on the action in the form of Hamilton's characteristic function. Corresponding to the propagator (Feynman's kernel) represented by Feynman's path integral, we introduce a novel object called the \emph{promotor} represented by the modified path integral.
By doing so, we are able to explore the power duality of the path integral formulation of quantum mechanics, at least formally, along the same line of thought as in the case of classical mechanics. We examine power-dual actions, path integrals, promotors, and Green functions. As is pointed out in reference \cite{IJ2021}, we find that the power duality is essentially a classical notion and that it breaks down upon angular quantization. By making a semiclassical Langer-like modification of angular quantization, we propose the partially-broken power-duality idea to utilize in carrying out path integration for certain potential systems.

In Section 2, we define the path integral for the promotor and clarify the relations of the promotor to the propagator and the Green function. Then we formulate the path integral in polar coordinates in a compact form and separate angular parts to extract the radial path integral for the power-duality consideration. We begin Section 3 by applying the space-time transformation to the radial path integral in Feynman's time slicing procedure. The full power-duality transformation $\Delta$ is also carefully applied to construct the power-dual pair of radial path integrals. Insofar as the radial promotor is characterized by a classical angular momentum $L$, the path integral for the promotor is power-dual form invariant.  As soon as the angular momentum is quantized, the power duality breaks down on the path integral for the promotor because $\mathfrak{L}$ cannot preserve the integral property of angular momentum quantum number $\ell$. This means that power duality is basically a classical notion. Nonetheless, there is a way to modify the classical notion so as to be useful as a tool in quantum mechanics.
While $\mathfrak{L}$ implies the equality $L_{a}=|\eta|^{-1} L_{b}$ which breaks down upon quantization,
we propose an \emph{ad hoc} replacement,
\begin{equation}\label{Ladhod}
  \mathfrak{L}_{\ell}$: \, $\ell + (D-2)/2 \rightarrow  |\eta | \{\ell + (D-2)/2 \}
\end{equation}
with $\ell \in \mathbb{N}_{0}$. We call this modified version of partially broken power-duality the \emph{quasi-power-duality}. Utilizing the modified version, we are able to obtain a quasi-power-dual formula for the Green functions which is similar to the one derived from the Schr\"odinger equation by Johnson \cite{Johnson}.
Section 4  deals with the power-dual pair of the hydrogen-like atom and the radial oscillator, and a family of confinement models as examples.
A brief summary is given in Section 5.

\section{Path Integral for the Promotor}\label{PIpromotor}

In this section, introducing an entity called \emph{promotor}, we study the path integral for the promotor. First we show how the promotor is related to the propagator (Feynman kernel) and the energy-dependent Green function. Then we present the polar coordinate formulation of the path integral. In particular, in order to be ready for investigating the power duality, we prepare the radial path integral by following Feynman's time slicing procedure.

\subsection{Propagator, Green Function and Promotor}

The time-dependent quantum state evolves as $|\psi(t'')\rangle = U(t'', t') |\psi(t')\rangle $ where $U(t'', t')$ is the time-evolution operator.
Since we are interested in a system whose Hamiltonian $H$ is not explicitly dependent on time $t$, the evolution operator can be given by  $U(t'', t')=\rme^{- \frac{\rmi}{\hbar}(t''-t')H}$.
In the $\mathbf{r}$-representation, the forward evolution equation for $t'' > t' $ takes the form,
\begin{equation}\label{evol}
\psi(\mathbf{r}'', t'') = \int \,\rmd \mathbf{r}' \,K(\mathbf{r}'', \mathbf{r}'; t'', t') \psi(\mathbf{r}', t').
\end{equation}
where $\psi(\mathbf{r}, t)=\langle \mathbf{r} | \psi(t)\rangle$ and
\begin{equation}\label{K1}
K(\mathbf{r}'', \mathbf{r}'; t'', t')= \langle \mathbf{r}'' |\rme^{- \frac{\rmi}{\hbar}(t''-t')H} | \mathbf{r}' \rangle \, \, \quad \, (t'' > t').
\end{equation}
As is defined by (\ref{K1}), $K(\mathbf{r}'', \mathbf{r}'; t'', t')$ is the transition amplitude of the system from position
$\mathbf{r}'$ at time $t'$ to position $\mathbf{r}''$ at time $t''$. In (\ref{evol}), it may be seen as the propagator that propagates quantum waves. It may as well be the kernel of the integral equation (\ref{evol}) and equivalently it is the retarded Green function ($t'' > t'$) for the time-dependent Schr\"odinger equation, $(H - \rmi\hbar \frac{\partial}{\partial t})\psi(\mathbf{r}, t)=0$.
As Feynman asserted,  the propagator can be evaluated by Feynman's path integral,
\begin{equation}\label{propa}
K(\mathbf{r}'', \mathbf{r}'; \tau)=\langle \mathbf{r}''| \rme^{-\frac{\rmi}{\hbar} H\tau} | \mathbf{r}'\rangle = \int_{_{\mathbf{r}'=\mathbf{r}(t')}}^{^{\mathbf{r}''=\mathbf{r}(t'')}} \mathcal{D}[\mathbf{r}(t)]\, \rme^{\frac{\rmi}{\hbar}S[\mathbf{r}(t)]}\,
\end{equation}
where $\tau =t''-t' > 0$ is the time interval of motion and $S[\mathbf{r}(t)]$ is the c-number action of the system in the form of Hamilton's principal function. Now on, the multi-faced function $K(\mathbf{r}'', \mathbf{r}'; t'', t')$ will be simply referred to as the \textit{propagator}.

The energy Green function $G(\mathbf{r}'', \mathbf{r}'; E)=\langle \mathbf{r}'' |(E - H)^{-1} |\mathbf{r}' \rangle$ for the stationary Schr\"odinger equation, $(H - E)\psi(\mathbf{r})=0$, is obtainable from the propagator (\ref{propa}) by a Laplace transformation as
\begin{equation}\label{Green1b}
G(\mathbf{r}'', \mathbf{r}'; E) =\frac{1}{i\hbar} \int _{0}^{\infty} \rmd\tau \, K(\mathbf{r}'', \mathbf{r}'; \tau)\, \rme^{\frac{\rmi}{\hbar}E\tau}\,, \quad {\rm Im \,}E > 0\,,
\end{equation}
by assuming that the energy parameter $E$ contains a small positive imaginary part ${\rm Im \,}E > 0$. Alternatively, we can express the energy Green function as
\begin{equation}\label{Green2a}
G(\mathbf{r}'', \mathbf{r}'; E)=\frac{1}{\rmi\hbar}\int _{0}^{\infty} \rmd\tau\, P(\mathbf{r}'', \mathbf{r}'; \tau)
\end{equation}
by introducing the object which we call the \textit{promotor} \cite{Inomata1986},
\begin{equation}\label{promo1}
P(\mathbf{r}'', \mathbf{r}'; t''-t')=\langle \mathbf{r}''| \rme^{-\frac{\rmi}{\hbar}(H-E)(t''-t')} | \mathbf{r}'\rangle\,, \qquad t'' > t'\,.
\end{equation}
The promotor $P(\mathbf{r}'', \mathbf{r}'; \tau)$ defined by (\ref{promo1}) does not propagate waves; it does not serve as a kernel. But it helps our duality argument as well as the calculation of the energy Green function (\ref{Green2a}).

Despite their difference in role, there is a similarity between the propagator and the promotor. When the propagator is for a system with the Hamiltonian $H$, the promotor may formally be viewed as the propagator for a system with a shifted Hamiltonian $\bar{H}=H - E$.
Therefore, it is obvious that the promotor can be evaluated by Feynman's path integral of the form,
\begin{equation}\label{promo2}
P(\mathbf{r}'', \mathbf{r}'; \tau) = \int_{_{\mathbf{r}'=\mathbf{r}(t')}}^{^{\mathbf{r}''=\mathbf{r}(t'')}}\mathcal{D}[\mathbf{r}(t)] \, \rme^{\frac{\rmi}{\hbar}W[\mathbf{r}(t)]} \,
\end{equation}
where Hamilton's principal function $S[\mathbf{r}(t)]$ in (\ref{propa}) is replaced by a modified action, $W[\mathbf{r}(t)]=S[\mathbf{r}(t)] + E\tau$, having the form of Hamilton's characteristic function.

In classical mechanics, as shown in reference \cite{IJ2021}, the radial action $W[r(t)]$ is dual-form invariant under the power-dual transformation $\Delta$. For the study of power duality in the path integral formulation of quantum mechanics, it is reasonable to choose the promotor in the modified path integral form (\ref{promo2}) rather than Feynman's original path integral (\ref{propa}) since the key constituent of the promotor path integral is $W[\mathbf{r}(t)]$ while Feynman's path integral depends on $S[\mathbf{r}(t)]$. Thus we investigate the power duality of the action $W[\mathbf{r}(t)]$ in the promotor.

If the action $W[\mathbf{r}(t)]$ is invariant under the set of operations, $\left\{\Delta, X(b, a)\right\}$, then the power duality of the promotor appears to be rather obvious. However, in applying $\Delta$ to the path integral (\ref{promo2}), we encounter a number of problems to be resolved. For instance, (i) the separation of the radial component from the angular parts, which is vital for the transformation of the radial variable, is not trivial in a path integral, (ii) the change of variables is not a straightforward matter in a path integral, and (iii) the action in a path integral is not necessarily defined along the classical paths; it covers all possible histories, so that the formal implementation of the non-integrable time transformation $\rmd\mathfrak{T}$ of (\ref{dT}) in a path integral is not warranted.
To resolve these problems, we have to look into the detailed structure of the path integral for the promotor. In fact, the path integral given in (\ref{promo2}) is still symbolic, whose implication has to be spelled out for computation.

It is clear from the properties of the evolution operator $U(t'', t')=\rme^{-\frac{\rmi}{\hbar}\bar{H}(t''-t')} $
 that the promotor defined by (\ref{promo1}) obeys, just as the propagator of (\ref{K1}) does,  the Kolmogorov-Chapman relation,
\begin{equation}\label{KC}
P(\mathbf{r}'', \mathbf{r}'; t''-t') = \int_{\mathbb{R}^{D}}\rmd^D\mathbf{r}\, P(\mathbf{r}'', \mathbf{r}; t''-t)\, P(\mathbf{r}, \mathbf{r}'; t-t')
\end{equation}
and the normalisation condition,
\begin{equation}\label{limit}
\lim_{\tau \rightarrow 0}\,P(\mathbf{r}'', \mathbf{r}'; \tau) = \delta^{(D)} (\mathbf{r}''-\mathbf{r}').
\end{equation}
Here, we have assumed that the space is Cartesian in $D$ dimensions. Replacing the $D$-dimensional integration of (\ref{KC}) by a star symbol $\ast$, we may interpret the Kolmogorov-Chapman relation (\ref{KC}) as the $\ast$  composition of a compatible pair of promotors; namely,
\begin{equation}\label{KC2}
P(\mathbf{r}'', \mathbf{r}'; t''-t') = P(\mathbf{r}'', \mathbf{r}; t''-t)\, \ast \, P(\mathbf{r}, \mathbf{r}'; t-t').
\end{equation}
We regard two promotors $P_{1}(t, t_{1})$ and $P_{2}(t_{2}, t)$ as a compatible pair since the final time of $P_{1}$ coincides with the initial time of $P_{2}$. Only a compatible pair of promotors are composable by the $\ast$ product.

Now, we follow Feynman's time slicing procedure \cite{Feynman1948} by dividing the time interval $\tau =t''-t' $ into $N$ short time subintervals $\tau_j$ in such a way that $\tau _{j} \sim \tau /N$ and $\tau = \sum_{j=1}^{N}\tau_{j}$, where $ j = 1,2,3, ... N$, $\tau_{j}=t_{j}-t_{j-1} > 0$, $t'=t_{0}$, and $t''=t_{N}$. We also let $\mathbf{r}_{j}=\mathbf{r}(t_{j})$, $\mathbf{r}'=\mathbf{r}(t')$ and $\mathbf{r}''=\mathbf{r}(t'')$. Repeating use of the Kolmogorov-Chapman relation (\ref{KC}), we may write the promotor as a multiple composition of short-time promotors $P(\mathbf{r}_{j}, \mathbf{r}_{j-1}; \tau_{j})$ defined for short time intervals $\tau_{j}$,
\begin{equation}\label{Psliced1}
  P(\mathbf{r}'', \mathbf{r}'; \tau)=\lim_{N \rightarrow \infty} \, \int \, \prod_{j=1}^{N-1} \rmd^{D} \mathbf{r}_{j}\prod_{j=1}^{N} P(\mathbf{r}_{j}, \mathbf{r}_{j-1}; \tau_{j})
\end{equation}
or
\begin{equation}\label{Psliced2}
  P(\mathbf{r}'', \mathbf{r}'; \tau)=\lim_{N \rightarrow \infty} \, \prod_{j=1}^{N} \ast P(\mathbf{r}_{j}, \mathbf{r}_{j-1}; \tau_{j}).
\end{equation}
In addition, the short time promotor is assumed to be
\begin{equation}\label{shortPro1}
P(\mathbf{r}_{j}, \mathbf{r}_{j-1}; \tau_{j})= \mathcal{N}_{j} \exp \left[\frac{\rmi}{\hbar} W(\mathbf{r}_{j}, \mathbf{r}_{j-1}; \tau_{j})\right]
\end{equation}
where $W(\mathbf{r}_{j}, \mathbf{r}_{j-1}; \tau_{j})$ is the short time action defined along a classical path, and $\mathcal{N}_{j}$ is a prefactor so chosen that the normalisation condition (\ref{limit}) is met.

The promotor given by (\ref{Psliced1}) and the short time promotor defined by (\ref{shortPro1}) constitute the time-slicing procedure of the path integral formally presented in (\ref{promo2}). This provides us a way to evaluate the path integral explicitly.

For the Lagrangian ${\cal L}(\mathbf{r}, \dot{\mathbf{r}})$ of a particle of mass $m$ moving in a central force potential $V(r)$, the short time action is given by
\begin{equation}\label{shortact1}
\fl
W(\mathbf{r}_{j}, \mathbf{r}_{j-1}; \tau_{j}) = \int_{t_{j-1}}^{t_{j}} \rmd t\left\{{\cal L}(\mathbf{r}, \dot{\mathbf{r}}) + E \right\} = \frac{m}{2\tau_{j}}(\Delta \mathbf{r}_{j})^{2} - V(r_{j})\tau_{j} + E\tau_{j}\,,
\end{equation}
where $\Delta \mathbf{r}_{j} = \mathbf{r}_{j} - \mathbf{r}_{j-1}$ and $r_{j}=|\mathbf{r}_{j}|$. The normalisation factor can be easily determined by using (\ref{limit}), (\ref{shortPro1}) and  (\ref{shortact1}); namely,
\begin{equation}\label{factor}
\mathcal{N}_{j} =\left[\frac{m}{2\pi \rmi \hbar \tau_{j}}\right]^{D/2}.
\end{equation}
In Feynman's construction \cite{Feynman1948,Feyn,Schu} of a path integral, the short time action is defined along a classical path.
Hence the infinitesimal time transformation $\rmd\mathfrak{T}$ is applicable to the short time promotor if it is properly replaced by an approximate short time transformation $\delta\mathfrak{T}$. The detail of this replacement will be discussed later.

\subsection{Path Integral in Polar Coordinates}\label{angularsubsection}

In order to perform $\mathfrak{R}$ of (\ref{powtrans}) that transform the radial variable, we have to express the promotor in spherical polar coordinates, say,
$(r, \theta^{(1)}, \theta^{(2)}, \cdots, \theta^{(D-1)})$, and extract its radial components out of the path integral. The polar coordinate formulation of the path integral can be found in the literature \cite{Peak,BJ87}. However, since it is not straightforward to move from the cartesian coordinates to the polar coordinates in a path integral as compared with the case of the Schr\"odinger equation, we shall briefly describe in the following the procedure to separate the radial promotor from the $D$ dimensional path integral. Details are presented in Appendix A.

Introducing the $D$ dimensional unit vector, $\mathbf{u}=\mathbf{r}/r \in \mathbb{R}^{D}$ indicating a point on the unit hypersphere $\mathbb{S}^{D-1}\subset \mathbb{R}^{D}$
we obtain for the short time promotor (\ref{shortPro1}) following partial wave expansion
\begin{equation}\label{shortPro3}
P(\mathbf{r}_{j}, \mathbf{r}_{j-1}; \tau_{j})=\sum_{\ell_{j}=0}^{\infty} P_{\ell_{j}}(r_{j}, r_{j-1}; \tau_{j})\,\mathcal{C}^{(D-2)/2}_{\ell_{j}}(\mathbf{u}_{j}\cdot \mathbf{u}_{j-1})\,,
\end{equation}
where $P_{\ell_{j}}(r_{j}, r_{j-1}; \tau_{j})$ is the short time radial promotor of the $\ell_{j}$-wave having the form,
\begin{equation}\label{shortR1}
\fl
P_{\ell_{j}}(r_{j}, r_{j-1}; \tau_{j}) = \frac{m}{\rmi\hbar \tau_{j}} \hat{r}_{j}^{-(D-2)}
\exp\left\{\frac{\rmi m}{2\hbar \tau_{j}} (r_{j}^{2} + r_{j-1}^{2}) - \frac{\rmi\tau_{j}}{\hbar}U_{j}\right\} I_{\ell_{j} + (D-2)/2}\left(\frac{m\hat{r}_{j}^{2}}{\rmi\hbar \tau_{j}}\right)
\end{equation}
with $\hat{r}_{j}^{2}=r_{j}r_{j-1}$ and $U_{j}=U(r_{j})$ where $U(r)$ is the modified potential $U(r)=V(r) - E$.
In the above $\mathcal{C}_{\ell}^{(D-2)/2}(\mathbf{u}\cdot \mathbf{u}')$ is the modified Gegenbauer polynomial defined in (\ref{AGmodif}) and obeys the orthonormality relation
\begin{equation}\label{GGG}
\int \rmd^{D-1}\Omega(\mathbf{u}) \, \mathcal{C}^{(D-2)/2}_{\ell''}(\mathbf{u}'' \cdot \mathbf{u})\, \mathcal{C}^{(D-2)/2}_{\ell'}(\mathbf{u} \cdot \mathbf{u}') = \delta_{\ell' \ell''} \, \mathcal{C}^{(D-2)/2}_{\ell'}(\mathbf{u}'' \cdot \mathbf{u}').
\end{equation}

Now, substituting the short time promotor of the form (\ref{shortPro3}) into (\ref{Psliced1}), we can easily carry out the angular integrations by utilizing the property (\ref{GGG}) successively to arrive at the finite time promotor expanded in partial wave series,
\begin{equation}\label{Pseries}
P(\mathbf{r}'', \mathbf{r}'; \tau)=\sum_{\ell=0}^{\infty} P_{\ell}(r'', r'; \tau)\,\mathcal{C}^{(D-2)/2}_{\ell}(\mathbf{u}'' \cdot \mathbf{u}'),
\end{equation}
with the radial promotor to be determined by path integration,
\begin{equation}\label{Rpathint1}
P_{\ell}(r'', r'; \tau) = \lim_{N \rightarrow \infty}\, \int \prod_{j=1}^{N-1} r_{j}^{D-1} \rmd r_{j}\, \prod_{j=1}^{N} P_{\ell}(r_{j}, r_{j-1}; \tau_{j}) .
\end{equation}

\subsection{Path Integral for the Radial Promotor}

The angular integrations have been completed. However, the radial integrations in (\ref{Rpathint1}) remain so far untouched.
In fact, the path integral for the radial propagator has been calculated only for a limited class of potentials including the free particle \cite{EdwardGulyaev}, the radial harmonic oscillator \cite{Peak}, and the hydrogen-like atom \cite{HoIno,Inomata1984,Steiner1984}. For a general power potential $V(r)= r^{n}$ with $n \neq -1, 0, 2,$ the radial path integration has not been carried out. Regardless of whether exact treatments are available or not, we are able to discuss the power duality in the radial path integral.

As justified in Appendix B we can utilise the asymptotic expression of the modified Bessel function for large arguments to approximate the short time radial promotor as follows (the symbol $\doteq$ indicates equality within this short time approximation)
\begin{equation}\label{radprom2}
P_{\ell}(r_{j}, r_{j-1}; \tau_{j})\doteq \left(\frac{m}{2\pi \rmi\hbar \tau_{j}}\right)^{1/2} \hat{r}_{j}^{-(D-1)} \, \exp\left\{\frac{\rmi}{\hbar }\, W(r_{j}, r_{j-1}; \tau_{j}) \right\}
\end{equation}
with the short time radial classical action,
\begin{equation}\label{shortwv}
W(r_{j}, r_{j-1}, \tau_{j})=\frac{m}{2\tau_{j}}(\Delta r_{j})^{2} - \frac{\left(L^{2} - \frac{1}{4}\right)\hbar^{2}}{2m \hat{r}_{j}^{2}} \tau_{j} - U(\hat{r}_{j})\tau_{j}
\end{equation}
where
\begin{equation}\label{angularm}
L=\ell + \frac{D-2}{2}, \, \quad \, \ell=0, 1, 2, \dots \,.
\end{equation}
Although quantum fluctuation does not only consist of classical paths,  in Feynman's time-slicing treatment, motion in a short time is assumed to be along a classical path. Hence the radial action (\ref{shortwv}) is considered as a classical action only when $\tau_{j}$ is very small, and
comparable to the classical action used in reference \cite{IJ2021} .  It must also be understood that  $V(r_{j}) \doteq V(\hat{r}_{j}) \doteq V(\bar{r}_{j})$ in the short time action because all terms of $O(\tau_{j}^{3/2})$ or higher can be ignored. Hence $U_{j}=U(r_{j}) \doteq U(\hat{r}_{j}) \doteq U(\bar{r}_{j})$, where $\hat{r}_{j} = \sqrt{r_{j} r_{j-1}}$ and $\bar{r}_{j}=(r_{j} + r_{j-1})/2$ are the geometric and arithmetic means, respectively.

Here we also note that the measure in the radial integral of (\ref{Rpathint1}) can be rearranged as
\begin{equation}\label{remeasure}
\prod_{j=1}^{N-1} r_{j}^{D-1} \rmd r_{j} = (r_{0} r_{N})^{-(D-1)/2} \prod_{j=1}^{N} \hat{r}_{j}^{D-1}\,\prod_{j=1}^{N-1} \rmd r_{j}.
\end{equation}
Substitution of the short time radial promotor (\ref{radprom2}) and the rearranged measure (\ref{remeasure}) into the path integral (\ref{Rpathint1}) enables us to write the $D$-dimensional radial promotor  (\ref{Rpathint1}) as
\begin{equation}\label{RpromF}
P_{\ell}(r'', r'; \tau)=(r'r'')^{-(D-1)/2} \mathcal{P}_{\ell}(r'', r'; \tau)
\end{equation}
with the one-dimensional radial promotor,
\begin{equation}\label{1Dpath}
\fl
\mathcal{P}_{\ell}(r'', r'; \tau) = \lim_{N \rightarrow \infty} \int \prod_{j=1}^{N-1} \rmd r_{j}\prod_{j=1}^{N}\left[\frac{m}{2\pi \rmi \hbar \tau_{j}}\right]^{1/2} \,\prod_{j=1}^{N}\exp \left\{\frac{\rmi}{\hbar}W(r_{j}, r_{j-1}; \tau_{j}) \right\}\,,
\end{equation}
where $W(r_{j}, r_{j-1}; \tau_{j})$ is the short time action given in (\ref{shortwv}). While the radial promotor $P_{\ell}(r'',r'; \tau)$ has been given by (\ref{radprom2}) as a radial path integral with the measure $\prod r_{j}^{D-1} \rmd r_{j}$, the radial promotor $\mathcal{P}_{\ell}(r'', r'; \tau)$ of (\ref{1Dpath}) is given by the one-dimensional radial path integral with the measure $\prod \rmd r_{j}$. The path integral representation for the one-dimensional promotor in (\ref{1Dpath}) is indeed based on the properties,
\begin{equation}\label{KC1d1}
\mathcal{P}_{\ell}(r'', r'; t'' -t') = \int_{\mathbb{R}_{+}} \rmd r\, \mathcal{P}_{\ell}(r'', r; t''-t) \mathcal{P}_{\ell}(r, r';t-t')
\end{equation}
and
\begin{equation}\label{1Dnorm}
\lim_{\tau \rightarrow 0} \mathcal{P}_{\ell}(r'', r'; \tau) = \delta (r'' - r').
\end{equation}
It is this one-dimensional radial path integral that we wish to study for the power duality.

The Green function (\ref{Green2a}) may as well be expanded in the partial wave series. Integrating both sides of (\ref{Pseries}) over $\tau$ yields
\begin{equation}\label{Green3a}
G(\mathbf{r}'', \mathbf{r}'; E)=\sum_{\ell=0}^{\infty} G_{\ell}(r'', r'; E)\,\mathcal{C}^{(D-2)/2}_{\ell}(\mathbf{u}'' \cdot \mathbf{u}')
\end{equation}
with the radial component defined in $D$-dimensional space,
\begin{equation}\label{Gradial1}
G_{\ell}(r'', r'; E)= \frac{1}{\rmi\hbar} \int _{0}^{\infty} \rmd\tau\, P_{\ell}(r'', r'; \tau) .
\end{equation}
The one-dimensional radial Green function corresponding to the one-dimensional radial promotor may be defined by
\begin{equation}\label{Gradial3}
\mathcal{G}_{\ell}(r'', r'; E)=\frac{1}{\rmi\hbar}\int_0^\infty \rmd\tau\,\mathcal{P}_{\ell}(r'', r'; \tau)
\end{equation}
which satisfies
\begin{equation}\label{Gradial2}
G_{\ell}(r'', r'; E)=(r'r'')^{-(D-1)/2} \mathcal{G}_{\ell}(r'', r'; E) .
\end{equation}

\section{Power Duality}
Here we wish to investigate the power duality on the radial path integral for the promotor. First, we consider the space and time transformations applied to a system with a central force potential. Then, restricting ourselves to a two-term power potential, we examine the transformation properties of the promotor and the Green function by applying the full power-duality transformation $\Delta$.

\subsection{Space and Time Transformations}
First we implement the transformation of radial variable, $\mathfrak{R}$,  and the time transformation, $\rmd \mathfrak{T}$, in the path integral for the one-dimensional radial promotor (\ref{1Dpath}).  Let system $A$ be the system described in terms of space time variables $(r, t)$ and system $B$ be the system in variables $(\rho, s)$.
In transforming a variable to another in the short time action, we have to exercise caution. For small $\varepsilon$, the following relations hold true \cite{IKG1992},
 \begin{equation}\label{gauss1}
 \int_{-\infty}^{\infty} \rmd x \,\exp\left\{ -\frac{\alpha}{\varepsilon }x^{2} + \beta x^{2} \right\} =   \int_{-\infty}^{\infty} \rmd x \,\exp\left\{ -\frac{\alpha}{\varepsilon}x^{2} + \frac{\beta}{2 \alpha}\varepsilon + O(\varepsilon^{2}) \right\},\,
 \end{equation}
\begin{equation}\label{gauss2}
  \int_{-\infty}^{\infty} \rmd x \,\exp\left\{ -\frac{\alpha}{\varepsilon}x^{2} + \frac{\gamma}{\varepsilon} x^{3} \right\} =   \int_{-\infty}^{\infty} \rmd x \,\exp\left\{ -\frac{\alpha}{\varepsilon}x^{2} + O(\varepsilon^{2}) \right\},\,
\end{equation}
\begin{equation}\label{gauss3}
  \int_{-\infty}^{\infty} \rmd x \,\exp\left\{ -\frac{\alpha}{\varepsilon}x^{2} + \frac{\delta}{\varepsilon} x^{4} \right\} =   \int_{-\infty}^{\infty} \rmd x \,\exp\left\{ -\frac{\alpha}{\varepsilon}x^{2} + \frac{3\delta}{4 \alpha}\varepsilon + O(\varepsilon^{2}) \right\},\,
\end{equation}
where $\alpha , \beta, \gamma$ and $\delta$ are constants, and ${\rm Re\,} \alpha > 0$.  Applying these relations to the short time action with $\varepsilon = \tau_{j}$ and $x = \Delta q_{j}$ for any real variable and recalling that the short time action neglects any terms of $O(\tau_{j}^{n})$ for $n > 1$, we obtain the following approximate relations,
\begin{equation}\label{relations}
\beta (\Delta q_{j})^{2} \doteq \frac{\beta}{2 \alpha} \tau_{j} \,, \qquad
\frac{\gamma}{\tau_{j}}(\Delta q_{j})^{3} \doteq 0 \,, \qquad
\frac{\delta}{\tau_{j}}(\Delta q_{j})^{4} \doteq \frac{3 \delta}{4 \alpha^{2}} \tau_{j}\,,
\end{equation}
which suggest that $(\Delta q_{j})^{2} \sim \tau_{j}$, and that the terms of $(\Delta q_{j})^{3}/\tau_{j}$ are negligible in the path integral and the terms of $(\Delta q_{j})^{4}/\tau_j$ are not ignorable. Naturally, all those terms containing $(\Delta q_{j})^{n}/\tau_j$ for $n > 4$ may be neglected.

Taking these points into account, we change each radial variable from $r_{j}$ to $\rho_{j}$ by the space transformation
\begin{equation}\label{ftrans}
\mathfrak{R}_{f}: \, \quad \, r \, \rightarrow \, \rho \, \quad \, \mbox{by} \, \quad \, r= f(\rho)
\end{equation}
where $f(\rho)$ is invertible and many times differentiable.  Let $f_{j}=f(\rho_{j})$ and $f'_{j}=f'(\rho_{j})$ with $f'=\rmd f/\rmd\rho$, etc. To treat two space points $\rho_{j}$ and $\rho_{j-1}$ symmetrically, we expand $f_{j}$ and $f_{j-1}$ as power series in $\Delta \rho_{j}$, and express $\Delta f_{j}= f_{j} -f_{j-1}$ in two possible ways;
\begin{equation}\label{expanf1}
\Delta f_{j} = f'_{j-1} (\Delta \rho_{j}) + \frac{1}{2!}f''_{j-1} (\Delta \rho_{j})^{2} + \frac{1}{3!} f'''_{j-1} (\Delta \rho_{j})^{3} + \cdots
\end{equation}
and
\begin{equation}\label{expanf2}
\Delta f_{j} = f'_{j} (\Delta \rho_{j}) - \frac{1}{2!}f''_{j} (\Delta \rho_{j})^{2} + \frac{1}{3!} f'''_{j} (\Delta \rho_{j})^{3} - \cdots
\end{equation}
Combining (\ref{expanf1}) and (\ref{expanf2}) we obtain
\begin{eqnarray}\label{expanf3}
\lefteqn{(\Delta f_{j})^{2} \doteq f'_{j} f'_{j-1} (\Delta \rho_{j})^{2} + \frac{1}{2!}\left\{ f'_{j} f''_{j-1} - f''_{j} f'_{j-1}\right\} (\Delta \rho_{j})^{3}} \nonumber  \\
    &  &  \, \quad + \frac{1}{3!}\left\{f'_{j} f'''_{j-1} + f'''_{j} f'_{j-1}\right\} (\Delta \rho_{j})^{4} - \frac{1}{2! 2!} f''_{j}f''_{j-1} (\Delta \rho _{j})^{4}.
\end{eqnarray}
In the above, we have ignored those terms containing $(\Delta \rho_{j})^{n}$ with $n > 4$.  However, if we hastily use the approximate relations of (\ref{relations}) at this point, we will miss an important contribution from the term of $(\Delta \rho_{j})^{3}$.
Since the leading term of (\ref{expanf3}) depends on the mean square value of $f'_{j}$, we have to proceed our calculation consistently by using the mean square values of derivatives,
$$
\hat{f}'_j=\sqrt{f'_jf'_{j-1}}\,\qquad \hat{f}''_j=\sqrt{f''_jf''_{j-1}}\,, \qquad \hat{f}'''_j=\sqrt{f'''_jf'''_{j-1}}\,.
$$
For conversion to the mean square values, we notice that
$$
\begin{array}{l}
\rho_{j}=\rho_{j-1}+\Delta \rho_{j}\,, \qquad f'_{j}=f'(\rho_{j})=f'_{j-1} + f''_{j-1}\Delta \rho_{j} + O((\Delta \rho_{j})^{2})\,,\\[2mm]
f'_{j-1}=f'_{j} - f''_{j}\Delta \rho_{j} + O((\Delta \rho_{j})^{2})\,, \qquad \hat{f}'_{j}=f'_{j} \{1 - \frac{1}{2}(f''_{j}/f'_{j})\Delta \rho_{j} + O((\Delta \rho_{j})^{2})\},
\end{array}
$$
etc., and that
\[
f'_{j}f''_{j-1}=f'_{j-1}f''_{j} + (f''_{j}f''_{j-1} - f'''_{j}f'_{j-1}) \Delta \rho_{j} + O((\Delta \rho_{j})^{2})
\]
and
\[
f^{(m)}_{j}f^{(n)}_{j-1}=\hat{f}^{(m)}_{j}\hat{f}^{(n)}_{j}\{1 - O(\Delta \rho_j)\}.
\]
Then we rewrite the second and third terms on the right hand side of (\ref{expanf3}) as
\begin{equation}\label{calcul1}
\frac{1}{2} \left\{f'_{j} f''_{j-1} - f''_{j} f'_{j-1}\right\} (\Delta \rho_{j})^{3} \doteq \frac{1}{2} \left\{(\hat{f}''_{j})^2 - \hat{f}'_{j}\hat{f}'''_{j}\right\} (\Delta \rho_{j})^{4}
\end{equation}
and
\begin{equation}\label{calcul2}
\frac{1}{6}(f'_{j}f'''_{j-1} + f'_{j-1} f'''_{j}) (\Delta \rho_{j})^{4} \doteq  \frac{1}{3} \hat{f}'_j\hat{f}'''_j (\Delta \rho_{j})^{4}.
\end{equation}
Surprising though it is, the term of $(\Delta \rho_{j})^{3}$ in (\ref{expanf3}) turns out to be of order $(\Delta \rho_{j})^{4}$ on the bases of mean square values.
Substituting the results of (\ref{calcul1}) and (\ref{calcul2}) into (\ref{expanf3}) yields
\begin{equation}\label{Deltar2}
(\Delta r_{j})^{2} = (\Delta f_{j})^{2} = (\hat{f'}_{j})^{2} (\Delta \rho_{j})^{2} + \frac{1}{6} (\hat{f}'_{j})^{2}\hat{\mathcal{S}}[\hat{f}_{j}] (\Delta \rho_{j})^{4}
\end{equation}
where
\begin{equation}\label{Sf}
\hat{\mathcal{S}}[\hat{f}_{j}] = \frac{\hat{f}'''_{j}}{\hat{f}'_{j}} - \frac{3}{2} \left(\frac{\hat{f}''_{j}}{\hat{f}'_{j}}\right)^{2}.
\end{equation}
Here $\hat{\mathcal{S}}[\hat{f}_{j}]$ is similar in form to but different in general from the Schwarz derivative $({\cal S}f)(\hat{\rho}_j)$.
However, as $\hat{f}^{(n)}_j =f^{(n)}(\hat{\rho}_{j})\{1+O(\Delta\rho_j)\}$, where $\hat{\rho}_{j}=\sqrt{\rho_{j}\rho_{j-1}}$, we may identify (\ref{Sf}) with the usual Schwarz derivative $({\cal S}f)(\rho)$. That is,
\begin{equation}\label{Sf2}
  \hat{\mathcal{S}}[\hat{f}_{j}] \doteq ({\cal S}f)(\hat{\rho}_{j})= \frac{f'''(\hat{\rho}_{j})}{f'(\hat{\rho}_{j})} - \frac{3}{2} \left(\frac{f''(\hat{\rho}_{j})}{f'(\hat{\rho}_{j})}\right)^{2}.
\end{equation}

By the change of variable $r_{j}=f(\rho_{j})$, the short time action $W^{(a)}(r_{j}, r_{j-1}; \tau_{j})$ in (\ref{shortwv}) is transformed to
\begin{equation}\label{WshortF2}
\fl
W^{(b)}(\rho_{j}, \rho_{j-1}; \tau_{j})=\frac{m}{2\tau_{j}}(\hat{f}'_{j})^{2} (\Delta \rho_{j})^{2}
 + \frac{\left(L^{2} - \frac{1}{4} \right)\hbar^{2}}{2m \hat{f}_{j}^{2}}\tau_{j} - V_{c}(\hat{f}_{j}) \tau_{j} - U(\hat{f}_{j})\tau_{j}\,
\end{equation}
where
\begin{equation}\label{correction1}
V_{c}(\hat{f}_{j}) \tau_{j}=-\frac{m}{12\tau_{j}} (\hat{f}'_{j})^{2} \hat{\mathcal{S}}[\hat{f}_{j}] (\Delta \rho_{j})^{4}.
\end{equation}
The extra potential (\ref{correction1}) is the non-ignorable contribution from the terms of $(\Delta \rho_{j})^{4}$ in (\ref{Deltar2}).
Bringing the action (\ref{WshortF2}) into the time-sliced promotor, we now employ the second approximate relation in (\ref{relations}) to convert $V_{c}(\hat{f}_{j})$ of (\ref{correction1}) into the form,
\begin{equation}\label{correction2}
V_{c}(\hat{f}_{j})=- \frac{\hat{\mathcal{S}}[\hat{f}_{j}] \hbar^{2}}{4m(\hat{f}'_{j})^{2}}
\end{equation}
which may be interpreted, being $V_c\sim \hbar^{2}$, as a quantum correction.

Correspondingly, the measure of the one-dimensional radial path integral (\ref{1Dpath}) undergoes the change,
\begin{equation}\label{measure3}
\prod_{j=1}^{N-1} \rmd r_{j}=(f'_{0}f'_{N})^{-1/2} \prod_{j=1}^{N} \hat{f}'_{j} \, \prod_{j=1}^{N-1} \rmd\rho_{j}.
\end{equation}



Next we consider the time transformation applied to the short time action  (\ref{WshortF2}) with the correction potential of (\ref{correction2}). The time transformation $\rmd{\mathfrak{T}}$ in Section \ref{Sec1} may be written in the form $\rmd \mathfrak{T}_{g}: \, \rmd t = g(\rho) \rmd s$. As the short time version of the infinitesimal time transformation $\rmd\mathfrak{T}_{g}$, we employ the invertible short time transformation
\begin{equation}\label{DeltaTg}
\delta \mathfrak{T}_{g}: \, \quad \, \tau_{j} = \hat{g}_{j} \sigma_{j}\,, \qquad j=1, 2, 3, ..., N,
\end{equation}
where $\tau_{j}=t_{j}- t_{j-1}$ and $\sigma_{j}=s_{j}-s_{j-1}$ are assumed to be small. Note that (\ref{DeltaTg}) does not stipulate how to
transform the finite time interval $\tau=\sum_{j}\tau_{j}=t''-t'$ to the corresponding interval $\sigma=\sum_{j}\sigma_{j}=s''-s'$.
Implementing $\delta \mathfrak{T}_{g}$ on the action (\ref{WshortF2}), we obtain the short time action,
\begin{equation}\label{WshortF3}
\fl
W^{(b)}(\rho_{j}, \rho_{j-1}; \sigma_{j})
=\frac{m}{2\sigma_{j}}\frac{(\hat{f}'_{j})^{2}}{\hat{g}_{j}}(\Delta \rho_{j})^{2} +\frac{\hat{g}_{j}\left(L^{2} - \frac{1}{4}\right)\hbar^{2}}{2m \hat{f}_{j}^{2}} \sigma_{j} - \hat{g}_{j}V_{c}(\hat{f}_{j}) \sigma_{j} - \hat{g}_{j}U(\hat{f}_{j})\sigma_{j}\,.
\end{equation}
Again, as has been argued in \cite{IJ2021}, in order to keep the form of the kinetic term in (\ref{WshortF3}) unchanged, we choose $\hat{g}_{j}=(\hat{f}'_{j})^{2}=f'(\rho_{j})f'(\rho_{j-1})$. Accordingly, (\ref{DeltaTg}) reads
\begin{equation}\label{DeltaT}
\delta \mathfrak{T}_{f}: \, \quad \, \tau_{j} = (\hat{f}'_{j})^{2} \sigma_{j}.
\end{equation}
Under $\delta \mathfrak{T}_{f}$, the action (\ref{WshortF3}) goes over to
\begin{equation}\label{WshortF4}
\fl
W^{(b)}(\rho_{j}, \rho_{j-1}; \sigma_{j})
 =\frac{m}{2\sigma_{j}}(\Delta \rho_{j})^{2} +\frac{(\hat{f}'_{j})^{2}\left(L^{2} - \frac{1}{4}\right)\hbar^{2}}{2m \hat{f}_{j}^{2}} \sigma_{j} - (\hat{f}'_{j})^{2}V_{c}(\hat{f}_{j}) \sigma_{j} - (\hat{f}'_{j})^{2}U(\hat{f}_{j})\sigma_{j}.\,
\end{equation}
This short time action differs in form from a classical action by the quantum correction term.
In a path integral, as is in the case of Schr\"odinger's equation, the change of variable gives rise to a correction term, see for example \cite{Junker1990}.

Under the operations of $\mathfrak{R}_{f}$ and $\delta \mathfrak{T}_{f}$, the measure in (\ref{1Dpath}), together with the normalisation factor, changes as
\begin{equation}\label{1Dmeasure}
\fl
\begin{array}{rcl}
\displaystyle
\prod_{j=1}^{N}\left[\frac{m}{2\pi \rmi \hbar \tau_{j}}\right]^{1/2} \prod_{j=1}^{N-1} \rmd r_{j}
&=& \displaystyle\prod_{j=1}^{N}\left[\frac{m}{2\pi \rmi \hbar \tau_{j}}\right]^{1/2}  \prod_{j=1}^{N-1} (f'_{j} \rmd\rho_{j}) \nonumber \\
&=& \displaystyle\left(f'_{0}f'_{N}\right)^{-1/2} \, \prod_{j=1}^{N}\left[\frac{m}{2\pi \rmi \hbar \tau_{j}}\right]^{1/2} \prod_{j=1}^{N}\hat{f}'_{j} \, \prod_{j=1}^{N} \rmd\rho_{j}  \nonumber \\
&=& \displaystyle\left(f'_{0}f'_{N}\right)^{-1/2}
\prod_{j=1}^{N}\left[\frac{m}{2\pi \rmi \hbar \sigma_{j}}\right]^{1/2} \,\prod_{j=1}^{N-1} \rmd\rho_{j}.
\end{array}\end{equation}

With the transformed action (\ref{WshortF4}) and the transformed measure (\ref{1Dmeasure}), the one dimensional radial path integral (\ref{1Dpath}) becomes
\begin{equation}\label{1Dpath2}
\mathcal{P}^{(a)}_{\ell}(r'', r'; \tau) =\tilde{\mathcal{P}}^{(b)}_{\ell}(\rho'', \rho'; \sigma) =\frac{1}{\sqrt{f'(\rho') f'(\rho'')}} \mathcal{P}^{(b)}_{\ell}(\rho'', \rho'; \sigma)
\end{equation}
where $f'(\rho')$ and $f'(\rho'')$ are the first derivatives of $f(\rho)$ with respect to $\rho$ at $\rho'$ and $\rho''$, respectively, and
\begin{equation}\label{1Dpath3}
\fl
\mathcal{P}^{(b)}_{\ell}(\rho'', \rho'; \sigma)
 = \lim_{N \rightarrow \infty} \int\prod_{j=1}^{N-1} \rmd\rho_{j}\,\prod_{j=1}^{N}
\left[\frac{m}{2\pi \rmi \hbar \sigma_{j}}\right]^{1/2}\,\prod_{j=1}^{N}\exp \left\{\frac{\rmi}{\hbar}W^{(b)}(\rho_{j}, \rho_{j-1}; \sigma_{j}) \right\}.
\end{equation}
In (\ref{1Dpath2}), $\tilde{\mathcal{P}}^{(b)}_{\ell}(\rho'', \rho'; \sigma)$ describes the form directly transformed, whereas $\mathcal{P}^{(b)}_{\ell}(\rho'', \rho'; \sigma)$ has a form similar to that of $\mathcal{P}^{(a)}_{\ell}(r'', r'; \tau)$ (i.e., symmetric under operation $X(b,a)$).
The short time action $W^{(b)}(\rho_{j}, \rho_{j-1}; \sigma_{j})$ on the right hand side of (\ref{1Dpath2}) is the one given by
(\ref{WshortF4}).  From (\ref{1Dnorm}) it is apparent that  $\tilde{\mathcal{P}}^{(b)}_{\ell}(\rho'', \rho'; \sigma)$ satisfies the normalisation condition,
\begin{equation}\label{1Dnewdelta}
\lim_{\sigma \rightarrow 0}\tilde{\mathcal{P}}^{(b)}_{\ell}(\rho'', \rho'; \sigma) = \frac{1}{\sqrt{f'(\rho') f'(\rho'')}} \delta( \rho'' - \rho') = \delta \left(f(\rho'') - f(\rho') \right),
\end{equation}
whereas $\mathcal{P}^{(b)}_{\ell}(\rho'', \rho'; \sigma)$ is normalised by
\begin{equation}\label{1Dnorm2}
\lim_{\sigma \rightarrow 0}\mathcal{P}^{(b)}_{\ell}(\rho'', \rho'; \sigma) = \delta( \rho'' - \rho').
\end{equation}
In this manner, the space and time transformation, $\mathfrak{R}_{f}$ plus $\delta \mathfrak{T}_{f}$, takes the one-dimensional radial path integral $\mathcal{P}^{(a)}_{\ell}(r'', r'; \tau_{j})$ of (\ref{1Dpath}) normalised by (\ref{1Dnorm}) to the one-dimensional radial path integral
$\tilde{\mathcal{P}}^{(b)}_{\ell}(\rho'', \rho'; \sigma)$ if normalised by (\ref{1Dnewdelta}) or
$\mathcal{P}^{(a)}_{\ell}(\rho'', \rho'; \sigma)$ of (\ref{1Dpath3}) if normalised by (\ref{1Dnorm2}).

While the time parameters $t$ and $s$ subjected to (\ref{DeltaT}) are those associated with path integration, the finte time intervals $\tau=t''-t'$ and $\sigma=s''-s'$  in the promotors, between which (\ref{DeltaT}) provides no transformation rule, are to be used as integration variables of Riemann integrals. The integration variables $\tau$ and $\sigma$ are both uniformly increasing in the range $(0, \infty)$.  Hence, even though  $\rmd t = g(\rho(s))\rmd s$ is not integrable, we may let $\rmd\tau=k \rmd\sigma$ where $k$ is a constant. To be compatible with the short time transformation (\ref{DeltaT}), we let $k=f'(\rho'')f'(\rho')$. Thus we have
\begin{equation}\label{dsigma}
\rmd\tau=f'(\rho'') f'(\rho') \rmd\sigma.
\end{equation}
Using the relations of (\ref{1Dpath2}) and (\ref{dsigma}) in (\ref{Gradial3}), we obtain
\begin{equation}\label{GandGbar}
\mathcal{G}^{(a)}_{\ell}(r'', r'; E_{a})=\sqrt{f'(\rho') f'(\rho'')} \, \mathcal{G}^{(b)}_{\ell}(\rho'', \rho'; E_{b}),
\end{equation}
where
\begin{equation}\label{Gbar}
\mathcal{G}^{(b)}_{\ell}(\rho'', \rho'; E_{b})=\frac{1}{\rmi\hbar}\int_0^\infty \rmd\sigma\,\mathcal{P}^{(b)}_{\ell}(\rho'', \rho'; \sigma),
\end{equation}
which is the one-dimensional radial Green function corresponding to the radial promotor (\ref{1Dpath3}) obtained by the space and time transformation.


\subsection{Completing Duality Operations}

Starting with the path integral for the promotor in $D$ dimensions, we have carried out the angular integrations in $D-1$ dimensions and obtained the one-dimensional radial path integral, $\mathcal{P}_{l}(r'', r'; \tau)$. Then, by performing the transformation of radial variable, $\mathfrak{R}_{f}$, and the short time transformation, $\delta \mathfrak{T}_{f}$, on the radial path integral, we have arrived at a similar one-dimensional radial path integral, $\mathcal{P}^{(b)}_{l}(\rho'', \rho'; \sigma)$, in (\ref{1Dpath3}).
However, the transformed path integral contains the action which is physically senseless until the transformation function $f(\rho)$ is explicitly spelled out. To be ready for discussing the power duality in the path integral formulation,
since the duality transformation for the path integral consists of the angular momentum transformation $\mathfrak{L}$, fixing a parameter of the space transformation $\mathfrak{C}$, and the energy-coupling interchange operation $\mathfrak{E}$, in addition to $\mathfrak{R}_{f}$ and $\delta \mathfrak{T}_{f}$, we have to perform the remaining operations, $\mathfrak{L}$, $\mathfrak{C}$ and $\mathfrak{E}$, and to shape up  $\mathcal{P}^{(b)}_{l}(\rho'', \rho'; \sigma)$ to a physically significant path integral.

The polar coordinate formulation of a path integral developed in the preceding sections is for a general central force potential $V(r)$.  What is needed for the present study is a power potential. To simplify our argument, we shall restrict ourselves to a two-term power potential of the form,
 \begin{equation}\label{2power}
 V(r)=V_a(r) =\lambda_{a} r^{a} + \lambda_{a'} r^{a'}\,,
 \end{equation}
 where $a, a' \neq 0$, $\lambda_{a}$ and $\lambda_{b}$ being real constants.

Let system $A$ be represented by the one-dimensional radial promotor $\mathcal{P}_{\ell}(r'', r'; \tau)$ of (\ref{1Dpath}) defined for the power potential (\ref{2power}). Introducing a prefactor $\varphi^{(a)}(r'r'')$ for symmetry adjustment, we consider the radial promotor of system $A$ as a path integral,
\begin{equation}\label{1DpathA}
\mathbf{P}^{(a)}(r'', r'; L_{a}; \tau) = \varphi^{(a)}(r'r'')\,\mathcal{P}^{(a)}(r'', r'; L_{a}; \tau)
\end{equation}
where
\begin{equation}\label{1DpathA2}
\fl
\mathcal{P}^{(a)}(r'', r'; L_{a}; \tau)= \lim_{N \rightarrow \infty} \int\prod_{j=1}^{N-1} \rmd r_{j} \prod_{j=1}^{N}\left[\frac{m}{2\pi \rmi \hbar \tau_{j}}\right]^{1/2} \,\prod_{j=1}^{N}\exp \left\{\frac{\rmi}{\hbar}W^{(a)}(r_{j}, r_{j-1}; \tau_{j}) \right\}
\end{equation}
with
\begin{equation}\label{shortwA}
W^{(a)}(r_{j}, r_{j-1}, \tau_{j})=\frac{m}{2\tau_{j}}(\Delta r_{j})^{2} - \frac{\left(L_a^{2} - \frac{1}{4}\right)\hbar^{2}}{2m \hat{r}_{j}^{2}} \tau_{j} - U^{(a)}(\hat{r}_{j})\tau_{j}
\end{equation}
and
\begin{equation}\label{UpowerA}
U^{(a)}(\hat{r}_{j}) = \lambda_{a} \hat{r}_{j}^{\,a}  + \lambda_{a'} \hat{r}_{j}^{\,a'} - E_{a}\,,  \qquad a \neq 0.
\end{equation}
Evidently $\mathbf{P}^{(a)}(r'', r'; L_{a}; \tau)$ satisfies
\begin{equation}\label{LimA}
\lim_{\tau \rightarrow 0} \mathbf{P}^{(a)}(r'', r'; L_{a}; \tau) = \varphi^{(a)}(r'r'') \delta(r'' - r').
\end{equation}
Here $L_{a}$ indicates the pre-quantized angular momentum of system $A$. In other words, $L_{a}=\ell_{a} + (D-1)/2$ with $\ell_{a} \in
\mathbb{N}_{0}$ is not assumed.
Let system $B$ be expressed in terms of $\mathcal{P}_{\ell}(\rho'', \rho'; \sigma)$ of (\ref{1Dpath3}) for the power potential (\ref{2power}) as
\begin{equation}\label{1DpathB}
\lefteqn{\mathbf{P}^{(b)}(\rho'', \rho'; L_{b}; \sigma) = \varphi^{(b)}(\rho'\rho'')\, \mathcal{P}^{(b)}(\rho'', \rho'; L_{b}; \sigma)},
\end{equation}
where
\begin{equation}\label{1DpathB2}
\fl
\mathcal{P}^{(b)}(\rho'', \rho'; L_{b}; \sigma)
 = \lim_{N \rightarrow \infty} \int \prod_{j=1}^{N-1} \rmd\rho_{j}\prod_{j=1}^{N}\left[\frac{m}{2\pi \rmi \hbar \sigma_{j}}\right]^{1/2} \,\prod_{j=1}^{N}\exp \left\{\frac{\rmi}{\hbar}W^{(b)}(\rho_{j}, \rho_{j-1}; \sigma_{j}) \right\}
\end{equation}
with
\begin{equation}\label{shortwB}
\fl W^{(b)}(\rho_{j}, \rho_{j-1}; \sigma_{j})
=\frac{m}{2\sigma_{j}}(\Delta \rho_{j})^{2} +\frac{(\hat{f}'_{j})^{2}\left(L_a^{2} - \frac{1}{4}\right)\hbar^{2}}{2m \hat{f}_{j}^{2}} \sigma_{j} - (\hat{f}'_{j})^{2}V_{c}(\hat{f}_{j}) \sigma_{j} - U^{(b)}(\hat{\rho}_{j}) \sigma_{j}
\end{equation}
and
\begin{equation}\label{UpowerB}
U^{(b)}(\hat{\rho}_{j}) = \hat{f}_{j}^{2}U^{(a)}(\hat{f}_{j}) =(\hat{f}'_{j})^{2}\left(\lambda_{a}(\hat{f}_{j})^{a}  + \lambda_{a'}(\hat{f}_{j})^{a'} - E_{a}\right) \,.
\end{equation}
In (\ref{1DpathB}), also, an undetermined prefactor $\varphi^{(b)}(\rho' \rho'')$ is inserted. Hence, $\mathbf{P}^{(b)}(\rho'', \rho'; L_{b}; \sigma)$ of (\ref{1DpathB}) obeys
\begin{equation}\label{LimB}
\lim_{\sigma \rightarrow 0}\mathbf{P}^{(b)}(\rho'', \rho'; L_{b}; \sigma) =\varphi^{(b)}(\rho' \rho'') \delta( \rho'' - \rho').
\end{equation}
Here $L_{b}$ signifies the angular momentum of system $B$, but in the action of (\ref{shortwB}) there is no quantity that can immediately be identified with the angular momentum of system $B$. It will be fixed only after the transformation function $f(\rho)$ is specified.

It is apparent that $W^{(a)}(r_{j}, r_{j-1}, \tau_{j})$ and $W^{(b)}(\rho_{j}, \rho_{j-1}; \sigma_{j})$ are the same short time action expressed in different sets of space and time variables.  They are equal but have different forms. Notice also that the measure of the path integral transforms as shown in the equality (\ref{1Dmeasure}) but is not form-invariant due to the presence of the factor $(f'(\rho') f'(\rho''))^{-1/2}$.
In general, $\mathbf{P}^{(a)}(r'', r'; L_{a}; \tau)$ does not transform to $\mathbf{P}^{(b)}(\rho'', \rho'; L_{b}; \sigma)$ under $\mathfrak{R}_{f}$ and $\delta \mathfrak{T}_{f}$. However, if there exist $\varphi^{(a)}(r'r'')$ and $\varphi^{(b)}(\rho' \rho'')$ such that
\begin{equation}\label{phiAphiB}
\frac{\varphi^{(a)}(r'r'')}{\varphi^{(b)}(\rho' \rho'')}=\sqrt{f'(\rho') f'(\rho'')},
\end{equation}
then system $A$ transforms to system $B$, that is,
\begin{equation}\label{PaPb}
\mathbf{P}^{(a)}(r'', r'; L_{a}; \tau)=\mathbf{P}^{(b)}(\rho'', \rho'; L_{b}; \sigma).
\end{equation}

In the above, we have used $L$ instead of $\ell$ to indicate the angular momentum. As will be seen later, the use of $\ell \in \mathbb{N}_{0}$ is misleading. The power duality transformation takes $L_{a}$ into $L_{b}$, but cannot take an integral $\ell_{a}$ into an integral $\ell_{b}$.
The subscript $\ell$ will be used only after the quasi-dual procedure is introduced.

The first step for performing the duality transformation is to choose a specific transformation function $f(\rho)$ of $\mathfrak{R}_{f}$. Let us specify $f(\rho)$ as a power function,
\begin{equation}\label{powerf}
\mathfrak{R}: \, \quad \, r=f(\rho) =C\rho^{\eta} \,, \qquad \eta > 0\,,
\end{equation}
where $C$ is a constant with a dimension of $\rho^{1- \eta }$.
For this function, $\hat{f}_{j}=C\hat{\rho}_{j}^{\eta}$ where $\hat{\rho}_{j}=\sqrt{\rho_{j}\rho_{j-1}}$ and
$\hat{f}'_{j}=\sqrt{f'_{j}f'_{j-1}} = C\eta \hat{\rho}_{j}^{\eta -1} = \rmd\hat{f}_{j}/\rmd\hat{\rho}_{j}$. Hence the time transformation associated with $\mathfrak{R}$ should be
\begin{equation}\label{DeltaTf}
\delta \mathfrak{T}_f: \, \quad \, \tau_{j} =C^{2}\eta^{2} \hat{\rho}_{j}^{2\eta -2} \sigma_{j}.
\end{equation}
It is also clear that the Schwarz derivative (\ref{Sf2}) takes the form,
\begin{equation}\label{Schwa2}
(\mathcal{S }f)(\hat{\rho}_{j}) = - \frac{\eta^{2} -1}{2 \hat{\rho}^{2}_{j}}.
\end{equation}
Accordingly, the quantum correction term in (\ref{WshortF4}) is given by
\begin{equation}\label{qcorrec}
(\hat{f}'_{j})^{2}V_{c}(\hat{f}_{j}) \sigma_{j} = \frac{(\eta^{2} - 1)\hbar^{2}}{8m \hat{\rho}_{j}^{2}}\sigma_{j}.
\end{equation}

Applying $\mathfrak{R}$ to the short time action (\ref{shortwB}) and combining the correction term (\ref{qcorrec}) to the angular momentum term in (\ref{shortwB}), we obtain
\begin{equation}\label{shortwB2}
W^{(b)}(\rho_{j}, \rho_{j-1}; \sigma_{j})=\frac{m}{2\sigma_{j}}(\Delta \rho_{j})^{2} +\frac{\left(\eta^{2}L_a^{2} - \frac{1}{4}\right)\hbar^{2}}{2m \rho_{j}^{2}} \sigma_{j} - U^{(b)}(\hat{\rho}_{j}) \sigma_{j}.
\end{equation}
with
\begin{equation}\label{UpowerB2}
U^{(b)}(\hat{\rho}_{j})=\lambda_{a} C^{2}\eta^{2}(\hat{\rho}_{j})^{a\eta + 2\eta - 2} + \lambda_{a'} C^{2}\eta^{2}(\hat{\rho}_{j})^{a'\eta + 2\eta - 2}
- E_{a}C^{2}\eta^{2} \hat{\rho}_{j}^{2\eta-2}.
\end{equation}

In the action (\ref{shortwB2}), it is apparent that the quantity corresponding to the angular momentum of system $B$ is $\eta L_{a}$ in the second term on the right-hand side. In fact, the angular momentum replacement operation contained in the duality transformation $\Delta$ dictates
\begin{equation}\label{angular2}
\mathfrak{L}: \, \quad \, L_b = \left|\eta \right| L_a,
\end{equation}
which we implement on the action (\ref{shortwB2}).

Next, on (\ref{UpowerB2}) we perform the two remaining operations that characterize the duality transformation by recalling
\begin{equation}\label{eta3}
\eta =- \frac{b}{a}= \frac{2}{a + 2} = \frac{b + 2}{2}\,, \qquad a \neq 0\quad\mbox{and}\quad a \neq  -2\,.
\end{equation}
The energy-coupling swap,
\begin{equation}\label{rename3}
\mathfrak{E}: \, \quad \, E_{b} = - \eta^{2}C^{a+2} \lambda_{a} \,, \qquad \lambda_{b} = - \eta^{2}C^{2} E_{a} \,,
\end{equation}
and the power transformation in the secondary potential,
\begin{equation}\label{second3}
\mathfrak{S}: \, \quad  \lambda_{b'}= \left(\frac{2}{a+2}\right)^{2} C^{a'+2}\lambda_{a'} \,, \qquad  b'=\frac{2(a'-a)}{a+2}\,.
\end{equation}
As a result, we reach the short time action,
\begin{equation}\label{shortwB3}
W^{(b)}(\rho_{j}, \rho_{j-1}; \sigma_{j})=\frac{m}{2\sigma_{j}}(\Delta \rho_{j})^{2} +\frac{\left(L_b^{2} - \frac{1}{4}\right)\hbar^{2}}{2m \rho_{j}^{2}} \sigma_{j} - U^{(b)}(\hat{\rho}_{j}) \sigma_{j}.
\end{equation}
with
\begin{equation}\label{UpowerB3}
U^{(b)}(\hat{\rho}_{j})=\lambda_{b} (\hat{\rho}_{j})^{b} + \lambda_{b'} (\hat{\rho}_{j})^{b'} - E_{b}.
\end{equation}
Just as the case of the finite time classical action, the short time action for a quantum particle remains form-invariant (symmetric under $X(a,b)$) under the full duality transformation $\Delta_{\delta}=\{\mathfrak{R},  \delta\mathfrak{T}, \mathfrak{L}, \mathfrak{E}, \mathfrak{S} \}$.

Now that $f(\rho)=C\rho^{\eta}$ is given by (\ref{powerf}) and $\eta$ by (\ref{eta3}), we have $f'(\rho) = (\eta/\rho)f(\rho)=-(b/a) (r/\rho)$ and
\begin{equation}\label{frfrho}
\sqrt{f'(\rho') f'(\rho'')}=\left| \frac{b}{a} \right| \frac{\sqrt{r'r''}}{\sqrt{\rho'\rho''}}.
\end{equation}
Hence we may choose the prefactors of the form,
\begin{equation}\label{PhiPhi}
\varphi^{(a)}(r'r'')=\frac{1}{|a|}\sqrt{r'r''}, \, \quad \,  \varphi^{(b)}(\rho' \rho'') =\frac{1}{|b|}\sqrt{\rho'\rho''}\,,
\end{equation}
to satisfy the condition (\ref{phiAphiB}).

With this choice of prefactors,  the path integral (\ref{1DpathA}) for the one-dimensional promotor of system $A$ transforms to that for the one-dimensional promotor of system $B$
under $\Delta_{\delta}$, that is,
\begin{equation}\label{1Dpath4}
\fl
\begin{array}{l}
\displaystyle
\mathbf{P}^{(b)}(\rho'', \rho'; L_{b}; \sigma) =  \frac{1}{|b|}\,\sqrt{\rho'\rho''}\\[4mm]
\displaystyle\qquad\qquad\times
\lim_{N \rightarrow \infty} \int \prod_{j=1}^{N-1} \rmd\rho_{j}\prod_{j=1}^{N}\left[\frac{m}{2\pi \rmi \hbar \sigma_{j}}\right]^{1/2} \,\prod_{j=1}^{N}\exp \left\{\frac{\rmi}{\hbar}W^{(b)}(\rho_{j}, \rho_{j-1}; \sigma_{j}) \right\} \,,
\end{array}
\end{equation}
where the short time action $W^{(b)}(\rho_{j}, \rho_{j-1}; \sigma_{j})$ is the one given by (\ref{shortwB3}) and supplemented by (\ref{UpowerB3}).

In (\ref{PaPb}), we have already seen that $\mathbf{P}^{(a)}(r'', r'; L_{a}; \tau)$ equals $\mathbf{P}^{(b)}(\rho'', \rho'; L_{b}; \sigma)$ via $\Delta_{\delta}$. If we let $\rho'=r_{a}$, $\rho'' =r_{b}$, $\tau=\tau_{a}$ and $\sigma=\tau_{b}$, then it is also apparent that $\mathbf{P}^{(b)}(\rho'', \rho'; L_{b}; \sigma)$ returns to $\mathbf{P}^{(a)}(r'', r'; L_{a}; \tau)$ under the parameter exchange operation
$X(a, b)$ insofar as they are expressed in the path integral form. However, once the path integration is carried out, the explicit exchange symmetry of the promotor will be lost.
Suppose that the path integration can be carried out for system $A$ and system $B$, and that the promotors for the two systems can be brought, respectively, to two closed form expressions depending on finite time intervals $\tau$ and $\sigma$. Since
the time-transformation involved in duality transformations are infinitesimal and not integrable, operation $\Delta_{\delta}$ does not stipulate a way to relate $\tau$ to $\sigma$. Hence, the $X(a,b)$ symmetry is no longer meaningful for the promotors given in closed form. In this regard, the exchange symmetry is not explicit. Nevertheless, we may say, though in a weaker sense,  the one-dimensional promotors,
$\mathbf{P}^{(a)}(r'', r'; L_{a}; \tau)$ and $\mathbf{P}^{(b)}(\rho'', \rho'; L_{b}; \sigma)$, are essentially power-dual to each other.

The one-dimensional radial Green functions of system $A$ and system $B$ are given, respectively, by (\ref{Gradial3}) and (\ref{Gbar}), satisfying (\ref{GandGbar}), namely,
\begin{equation}\label{GandGbar2}
\mathcal{G}^{(a)}(r'', r'; L_{a}; E_{a})=\sqrt{f'(\rho') f'(\rho'')} \, \mathcal{G}^{(b)}(\rho'', \rho'; L_{b}; E_{b}).
\end{equation}
In a manner parallel to the symmetrized (bold-faced) promotors, we define the  symmetrized Green functions,
\begin{equation}\label{GradialA3}
\mathbf{G}^{(a)}(r'', r'; L_{a}; E_{a})=[\varphi^{(a)}(r'r'')]^{-1}\, \mathcal{G}^{(a)}(r'', r'; L_{a}; E_{a}),
\end{equation}
and
\begin{equation}\label{GradialB3}
\mathbf{G}^{(b)}(\rho'', \rho'; L_{b}; E_{b})= [\varphi^{(b)}(\rho'\rho'')]^{-1} \,  \mathcal{G}^{(b)}(\rho'', \rho'; L_{b}; E_{b}).
\end{equation}
Then we have
\begin{equation}\label{GaGb}
\mathbf{G}^{(a)}(r'', r'; L_{a}; E_{a})=\mathbf{G}^{(b)}(\rho'', \rho'; L_{b}; E_{b}),
\end{equation}
which means that $\mathbf{G}^{(a)}(r'', r'; L_{a}; E_{a})$ is brought to $\mathbf{G}^{(b)}(\rho'', \rho'; L_{b}; E_{b})$ by $\Delta_{\delta}$.
To carry out the time integration in (\ref{Gbar}), the promotor must be in a closed form expression. Hence the resultant Green functions may not generally be symmetric under $X(a, b)$. Again, in a weaker sense, we may say that
the Green functions (\ref{GradialA3}) and (\ref{GradialB3}) are power-dual to each other.

It is important, however, to recognize that the implicit power-dual symmetry possessed by the radial promotors and the corresponding Green functions breaks down as soon as the angular momentum $L$ is quantized. Suppose each of $L_{a}$ and $L_{b}$ is given by (\ref{angularm}), that is,
\begin{equation}\label{angmoAB}
L_{a}=\ell_{a} + (D_{a}-2)/2, \, \, \quad \, L_{b}=\ell_{b} + (D_{b}-2)/2.
\end{equation}
Then the angular duality operation, $\mathfrak{L}$: $L_{b}=\eta L_{a}$, should imply
\begin{equation}\label{ellABtrans}
\ell_{b} +(D_{b}-2)/2 =\eta \ell_{a} + \eta (D_{a}-2)/2,
\end{equation}
which does not warrant that $\ell_{b}$ will be integral when $\ell_{a}$ is an integer since $\eta$ can be fractional.
Furthermore, if the space dimension is assumed to be a positive integer, it is difficult for both $\ell_{a}$ and $\ell_{b}$ to be integral simultaneously.  An exceptional example is the case where $D_{a}=3$, $D_{b}=2$ and $\eta=2$. In this case, $\ell_{a}=0, 1, 2, ...$ corresponds to $\ell_{b}=1, 3, 5, ...$. As has been mentioned in \cite{IJ2021}, this situation occurs in the correspondence between the full three-dimensional Coulomb problem and the odd-half of the two-dimensional harmonic oscillator at the spectrum level. Such a correspondence between two different dimensions is beyond the context of the power duality argument.

Since we have been considering the power duality between two systems in the same dimension, we let $D_{a}=D_{b}$. Then (\ref{ellABtrans}) is simplified as
\begin{equation}\label{ellABtrans2}
\ell_{b} =\eta \ell_{a} + (\eta -1)(D-2)/2,
\end{equation}
with which the situation still remains unimproved.

To construct any $D$ dimensional full promotor $P(\bf{r}'', \bf{r}'; \tau)$ expressed in spherical polar coordinates by the partial wave series expansion (\ref{Pseries}), we need a complete set of partial wave components $\mathcal{P}_{\ell}(r'', r'; \tau)$ with $\ell=0,1,2,...$. Hence the angular operation (\ref{ellABtrans}) must bring us the correspondence between $\ell_{a}=0, 1, 2, ...$ and $\ell_{b}=0, 1, 2, ...$, but fails in general. Therefore, even if two $L$ dependent radial promotors, $\mathcal{P}^{(a)}(r'', r'; L_{a}; \tau)$ and $\mathcal{P}^{(b)}(\rho'', \rho'; L_{b}; \sigma)$, are power-dual to each other, the corresponding partial wave promotors $\mathcal{P}_{\ell_{a}}(r'', r'; \tau)$ with $\ell_{a}=0,1,2,...$ and $\mathcal{P}_{\ell_{b}}(\rho'', \rho'; \sigma)$ with $\ell_{b}=0,1,2,...$ are not exactly a power-dual pair. Consequently, the corresponding full promotors, $P^{(a)}(\bf{r}'', \bf{r}'; \tau)$ and $P^{(b)}(\bm{\rho}'', \bm{\rho}'; \sigma)$, are not power-dually symmetric.
This leads us to a conclusion that the path integral formulation of quantum mechanics as a whole does not possess the power duality symmetry.

Having said so, we are still able to utilize the basic idea of power duality for path integrals by making a semi-classical like interpretation of operation $\mathfrak{L}: \,L_{b}=\eta L_{a}$. First, we consider $\mathfrak{L}$ as a classical operation by giving up the transformation of the form  (\ref{ellABtrans}). After $\mathfrak{L}$ is applied, we make an \emph{ad hoc replacements},
\begin{equation}\label{ellAellB}
\mathfrak{L}_{q}: \,L_{a}=\ell  + (D-2)/2, \, \, \quad \, L_{b}=\ell + (D-2)/2, \, \, \quad \, (\ell \in \mathbb{N}_{0}).
\end{equation}
As a result of (\ref{ellAellB}),  the actions (\ref{shortwA}) and (\ref{shortwB}) are expressed, respectively, by
\begin{equation}\label{shortwA2}
\fl\begin{array}{l}
W^{(a)}_{\ell}(r_{j}, r_{j-1}, \tau_{j}) = \\
\displaystyle\qquad\qquad
\frac{m}{2\tau_{j}}(\Delta r_{j})^{2} - \frac{\left(\ell+(D-3)/2\right)\left(\ell+(D-1)/2\right)\hbar^{2}}{2m \hat{r}_{j}^{2}} \tau_{j} - \left(\lambda_{a}r^{a} - E_{a}\right)\tau_{j},
\end{array}\end{equation}
and
\begin{equation}\label{shortwB4}
\fl\begin{array}{l}
W^{(b)}_{\ell}(\rho_{j}, \rho_{j-1}; \sigma_{j})=\\
\displaystyle\qquad\qquad
\frac{m}{2\sigma_{j}}(\Delta \rho_{j})^{2} +\frac{\left(\ell+(D-3)/2\right)\left(\ell+(D-1)/2\right)
\hbar^{2}}{2m \hat{\rho}_{j}^{2}} \sigma_{j} - \left(\lambda_{b}\rho^{b} - E_{b}\right) \sigma_{j}.
\end{array}
\end{equation}
Although the $L_{a}$ dependent action $W^{(a)}(r_{j}, r_{j-1}; \tau_{j})$ transforms into the $L_{b}$ dependent action  $W^{(b)}(\rho_{j}, \rho_{j-1}; \sigma_{j})$ under the power duality transformation $\Delta$, the $\ell$ dependent
$W^{(a)}_{\ell}(r_{j}, r_{j-1}, \tau_{j})$ does not transform to the $\ell$ dependent $W^{(b)}_{\ell}(\rho_{j}, \rho_{j-1}; \sigma_{j})$ under $\Delta$. Therefore,
$W^{(a)}_{\ell}(r_{j}, r_{j-1}, \tau_{j})$ and $W^{(b)}_{\ell}(\rho_{j}, \rho_{j-1}; \sigma_{j})$ are no longer power dual to each other. Nevertheless, it is true that $W^{(b)}_{\ell}(\rho_{j}, \rho_{j-1}; \sigma_{j})$ goes back to $W^{(a)}_{\ell}(r_{j}, r_{j-1}, \tau_{j})$ by $X(a, b)$.
If we define the  \emph{quasi-power-dual} transformation by $\Delta_{q}=\{\Delta, \mathfrak{L}_{q}\}$, then
\begin{equation}\label{waqwb}
W^{(a)}_{\ell}(r_{j}, r_{j-1}, \tau_{j}) \simeq W^{(b)}_{\ell}(\rho_{j}, \rho_{j-1}; \sigma_{j}),
\end{equation}
where $\simeq$ signifies the equality under $\Delta_{q}$. Thus, these $\ell$ dependent short time actions are dual to each other with respect to $\Delta_{q}$, or quasi-power-dual to each other. Correspondingly, if we define the one-dimensional promotors for the $\ell$-th wave by
\begin{equation}\label{PromAell}
\mathbf{P}^{(a)}_{\ell}(r'', r'; \tau) = \left.\mathbf{P}^{(a)}(r'', r';L_a; \tau)\right|_{L_{a}=\ell+(D-2)/2} \,, \qquad\ell \in \mathbb{N}_{0}\,,
\end{equation}
and
\begin{equation}\label{PromBell}
\mathbf{P}^{(b)}_{\ell}(\rho'', \rho';  \sigma) = \left.\mathbf{P}^{(b)}(\rho'', \rho';L_b; \sigma)\right|_{L_{b}=\ell+(D-2)/2} \,,\qquad \ell \in \mathbb{N}_{0},
\end{equation}
then we may say that $\mathbf{P}^{(a)}_{\ell}(r'', r'; \tau)$ and $\mathbf{P}^{(b)}_{\ell}(\rho'', \rho';  \sigma)$ are quasi-dual to each other, and write as
\begin{equation}\label{PellqAB}
\mathbf{P}^{(a)}_{\ell}(r'', r'; \tau) \simeq \mathbf{P}^{(b)}_{\ell}(\rho'', \rho';  \sigma).
\end{equation}

The $\ell$ dependent Green functions can be constructed by using the $\ell \in \mathbb{N}_{0}$ dependent promotors, (\ref{PromAell}) and (\ref{PromBell});  that is, ,
\begin{equation}\label{Gaell}
\mathbf{G}^{(a)}_{\ell}(r'', r';  E_{a})=\left.\mathbf{G}^{(a)}(r'', r'; L_{a}; E_{a})\right|_{L_{a}=\ell + (D-2)/2}\,,
\end{equation}
and
\begin{equation}\label{Gbell}
\mathbf{G}^{(b)}_{\ell}(\rho'', \rho';  E_{b})=\left.\mathbf{G}^{(b)}(\rho'', \rho'; L_{b}; E_{b})\right|_{L_{b}=\ell + (D-2)/2}\,,
\end{equation}
respectively. They are also quasi-power-dual to each other,
\begin{equation}\label{GellqAB}
\mathbf{G}^{(a)}_{\ell}(r'', r';  E_{a}) \simeq \mathbf{G}^{(b)}_{\ell}(\rho'', \rho';  E_{b}).
\end{equation}

Utilizing the quasi-dual relation (\ref{GellqAB}), we are able to determine the $\ell$ dependent one-dimensional Green function of system $A$ if we know the form of the $\ell$ dependent one-dimensional Green function of system $B$. Namely, we obtain
\begin{equation}\label{GbGa}
\fl
\begin{array}{l}
\mathbf{G}^{(a)}_{\ell}\left(r'', r'; \ell + \frac{D-2}{2}; E_{a}; \lambda_{a};\lambda_{a^{(i)}}\right) = \\
\qquad\mathbf{G}^{(b)}_{\ell}\left((r'/C)^{1/\eta}, (r''/C)^{1/\eta}; \eta(\ell + \frac{D-2}{2}); -\eta^{2} C^{a+2}\lambda_{a}; - \eta^{2} C^{2} E_{a}; \lambda_{b^{(i)}}\right) \,,
\end{array}
\end{equation}
where $\ell \in \mathbb{N}_{0}$, $\lambda_{b^{(i)}} = \left(\frac{2}{a+2}\right)^{2} C^{a^{(i)}+2}\lambda_{a^{(i)}}$ and $\eta=-b/a$ with $a \neq 0$.
Here we added two more parameters for the coupling constants to reflect the energy-coupling swapping (\ref{rename}) and the change in the secondary potential (\ref{second}).
This formula is similar to the relation obtained for the Green function satisfying the Schr\"odinger equation by Johnson \cite{Johnson}.

Note that this type of trick is not applicable in finding a closed form expression of the propagator or the promotor because the finite time transformation rule lacks in $\Delta_\delta$.

\section{Coulomb-Hooke Pair and Confinement Model}
Since the exactly path integrable systems are very limited, we select the Coulomb-Hooke dual pair and a family of confinement models as examples of solvable dual pairs.

\subsection{Example 1:  Coulomb-Hooke dual pair $(a, b) = (-1,  2)$}
Let system $A$ be the radial part of  the hydrogen-like atom in $3$ dimensional space and system $B$ be a $3$ dimensional radial harmonic oscillator. So we have a dual pair $(a,b) = (-1, 2)$ with $\lambda_{a}=-Ze^{2}$, $\lambda_{b}=\frac{1}{2}m\omega^{2}$, $E_{a}= - Z^{2}e^{4}m/[2\hbar ^{2}(n+ \ell + 1)^{2}]$ and $E_{b}=\hbar \omega (2n + \ell + 3/2)$, where $n, \ell \in \mathbb{N}_{0}$.

The harmonic oscillator is a typical example for which Feynman's path integral can be exactly evaluated. Feynman \cite{Feynman1948} derived the propagator for the oscillator from the path integral in cartesian coordinates. For the radial oscillator, Peak and Inomata \cite{Peak} obtained the propagator from the radial path integral. On the other hand, the propagator for the Coulomb system had been considered difficult to construct in closed form by path integration or otherwise. As for the Coulomb Green function, Hostler \cite{Hostler} obtained a closed form expression as well as a series expansion in polar coordinates by using the solutions of the Schr\"odinger equation.

Despite the fact that the hydrogen atom solution symbolized the success of the Schr\"odinger equation,
Feynman's path integral with the Coulomb potential cannot be solved.
By applying formally the Kustaanheimo-Stiefel transformation \cite{KS1965} to the path integral in the Hamiltonian formulation, Duru and Kleinert \cite{DuruKleinert} succeeded to derive the Green function for the hydrogen atom in the momentum representation. Similarly. though explicitly, implementing the Kustaanheimo-Stiefel transformation in Feynman's path integral in the Lagrangian formulation, Ho and Inomata \cite{HoIno} obtained the Green function for the hydrogen atom in polar coordinates.
Soon after, Inomata \cite{Inomata1984} and Steiner \cite{Steiner1984} independently found a simplified way to calculate the radial Green function for the Coulomb problem by applying a space-time transformation to the radial path integral. Now, recognizing that many previously employed procedures are covered by the unified framework of power-duality.

As has been seen in Section 3, the short time radial action $W(r_{j}, r_{j-1}, \tau_{j})$ is dual-form invariant under $\Delta_\delta$. For the dual pair $(a, b)=(-1, 2)$, we have $\eta=-b/a=2$. In this case, $a'=b'=0$.  The corresponding duality transformation $\Delta_\delta$ consists of operations,
\begin{equation}\label{powt1}
\mathfrak{R}: \, \quad \, r=f(\rho) =C\rho^{2},
\end{equation}
\begin{equation}\label{dT1}
\delta \mathfrak{T}: \quad  \tau_{j} =4 C^{2} \rho^{2} \sigma_{j},
\end{equation}
\begin{equation}\label{angtrans1}
\mathfrak{L}: \, \quad \, L_{b} = 2 L_{a},
\end{equation}
\begin{equation}\label{rename1}
\mathfrak{E}: \, \quad \, E_{b} = - 4C \lambda_{a}, \, \quad \, \lambda_{b} = - 4C^{2} E_{a}.
\end{equation}
Under $\Delta_\delta$ with $\eta = 2$,  the short time radial action for the hydrogen-like atom,
\begin{equation}\label{Coulact}
W_{a}(r_{j}, r_{j-1}, \tau_{j})=\frac{m}{2\tau_{j}}(\Delta r_{j})^{2}
- \frac{\left(L_{a}^{2} - \frac{1}{4}\right) \hbar^{2}}{2m \hat{r}_{j}^{2}} \tau_{j} + \frac{Ze^{2}}{\hat{r}_{j}}\tau_{j} + E_{a}\tau_{j}\,,
\end{equation}
transforms into the one for the radial oscillator,
\begin{equation}\label{Hookeact}
W_{b}(\rho_{j}, \rho_{j-1}; \sigma_{j})
=\frac{m}{2\sigma_{j}}(\Delta \rho_{j})^{2} +\frac{\left(L_{b}^{2} - \frac{1}{4}\right)\hbar^{2}}{2m \hat{\rho}_{j}^{2}} \sigma_{j}
-\frac{1}{2}m\omega^{2} \hat{\rho}_{j}^{2} \sigma_{j} + E_{b} \sigma_{j}
\end{equation}
with
for which path integration has been explicitly carried out. Therefore, using the path integration result for the radial harmonic oscillator (the oscillator or Osc in short), we are able to construct at least the Green function for the hydrogen-like atom (the Coulomb system or Coul in short) via the quasi-dual formula (\ref{GellqAB}).

The propagator derived by Peak and Inomata \cite{Peak} for the radial oscillator when $L_{b}=\ell + 1/2 $, $\ell\in \mathbb{N}_{0}$, is
\begin{equation}\label{OsciK}
\fl
\begin{array}{rcl}
K^{(Osc)}_{\ell}(\rho'', \rho'; \sigma) &=& \displaystyle\frac{m\omega}{\rmi\hbar \sqrt{\rho'\rho''}\,\sin(\omega\sigma)}\\[4mm]
&&\times \displaystyle\exp\left\{\frac{\rmi m\omega}{2\hbar \tan(\omega\sigma)}\left(\rho'^{2} + \rho''^{2} \right) \right\}
\, I_{\ell + \frac{1}{2}} \left(\frac{m\omega \rho'\rho''}{\rmi \hbar \sin(\omega \sigma)}\right)
\end{array}\end{equation}
which,  as is clear from (\ref{BBessel}) for $\sigma$ small, satisfies the condition,
\begin{equation}\label{OsciNC}
\lim_{\sigma \rightarrow 0}K^{(Osc)}_{\ell}(\rho'', \rho'; \sigma) = \frac{1}{\rho'\rho''}\delta(\rho''- \rho').
\end{equation}
The corresponding one-dimensional promotor is
\begin{equation}\label{OsciP}
\fl
\begin{array}{l}
\mathcal{P}^{(Osc)}_{\ell}(\rho'', \rho'; \sigma)  =  \rho'\rho'' K^{(Osc)}_{\ell}(\rho'', \rho'; \sigma)\rme^{\frac{\rmi}{\hbar}E_{b}\sigma}\\[4mm]
 \qquad =  \displaystyle\frac{m\omega \sqrt{\rho'\rho''}}{ \rmi\hbar\,\sin(\omega\sigma)}
\exp\left\{\frac{\rmi m\omega}{2\hbar \tan(\omega\sigma)}\left(\rho'^{2} + \rho''^{2} \right)\right\}\rme^{\frac{\rmi}{\hbar}E_{b}\sigma}
\, I_{\ell + \frac{1}{2}} \left(\frac{m\omega \rho'\rho''}{\rmi \hbar \sin(\omega \sigma)}\right),
\end{array}
\end{equation}
which is normalised as
\begin{equation}\label{OsciPn}
\lim_{\sigma \rightarrow 0}\mathcal{P}^{(Osc)}_{\ell}(\rho'', \rho'; \sigma)=\delta(\rho'' - \rho').
\end{equation}

Although we know that
$\mathbf{P}^{(Osc)}(\rho'', \rho'; L_{b}; \sigma)=\varphi^{(b)}(\rho'\rho'')\mathcal{P}^{(Osc)}_{\ell}(\rho'', \rho'; \sigma)
$ has its dual partner $\mathbf{P}^{(Coul)}(r'', r'; L_{a} ; \tau)$, neither the propagator (\ref{OsciK}) nor the promotor (\ref{OsciP}) is directly helpful in predicting the counterpart of the partner system because the transformation rule of time intervals $\sigma \rightarrow \tau $ lacks in $\Delta_\delta$. Since the Green function is time-independent,  we attempt to reach the Coulomb system from the Green function for the radial oscillator.

The Green function for Osc is in principle obtainable by integrating the promotor (\ref{OsciP}) over $\sigma$ as shown in (\ref{Gradial1}),  that is,
\begin{equation}\label{gosc1}
\mathcal{G}^{(Osc)}_{\ell}(\rho'', \rho'; E_{b})= \frac{1}{\rmi\hbar}
\int _{0}^{\infty} \rmd\sigma\, \mathcal{P}^{(Osc)}_{\ell}(\rho'', \rho';  \sigma)
\end{equation}
which is nonetheless problematic. As $\mathcal{P}^{(Osc)}_{\ell}(\rho'', \rho';  \sigma)$ of (\ref{OsciP}) diverges when $\omega \sigma = n\pi \, (n\in \mathbb{N}_{0})$, the $\sigma$ integration from $0$ to $\infty$ cannot be achieved. To circumvent this problem, we treat the integration variable as a complex number by letting $z=\omega \sigma$ and the promotor as a complex function of $z$ by letting $F(z)=(1/\rmi\hbar) \mathcal{P}^{(Osc)}_{\ell}(\rho'', \rho'; z/\omega)$. Since $F(z)$ has simple poles at $z=z_{n}=n\pi \, (n \in \mathbb{N}_{0})$ and is analytic in the lower right quadrant ($ \mathrm{Re\,} z \geq 0, \mathrm{Im\,} z < 0$) of the $z$-plane, we consider the integral along a closed contour $C$ satisfying Cauchy's theorem, $\oint_{C} \rmd z F(z) = 0$.
Writing $z=\alpha + \rmi\beta = q \rme^{\rmi\theta}$ where $q=\sqrt{\alpha^{2} + \beta^{2}}$, $\tan \theta= \alpha/\beta$ and $\alpha, \beta, \in \mathbb{R}$,  we choose $C$ to be consisting of three parts(see also Figure 1):
$$
\begin{array}{rcl}
 {\cal C}_{1} &=& \{ {\rm Re\,}z =\alpha_{n}\, |\,  z_{n}+\varepsilon \leq \alpha_{n} \leq z_{n+1}- \varepsilon, \, \, \, 0 <\varepsilon < \frac{\pi}{2}, \, n \in \mathbb{N}_{0}\, \}\\
       & & \cup\, \{ z=z_{n} + \varepsilon \rme^{\rmi\phi_{n}}\, |\, -\pi/2 \leq \phi_{0} \leq 0, \,\,-\pi \leq \phi_{n} \leq 0, \, \, n \in \mathbb{N} \}\\
       & & {\rm from~} \alpha=0 {\rm ~to~} \alpha=q_{1} {\rm ~along~ the~ real~ line~ of~} z;\\[2mm]
{\cal C}_{2}  &=& \{ z=q_{1}e^{i\theta} \,| -\pi/2 \leq \theta \leq 0 \, \} {\rm ~from~} \theta=0 {\rm ~to~} \theta = -\pi/2\, {\rm ~(clockwise); \, and} \\[2mm]
{\cal C}_{3}  &=&\{- q_{1} \leq \beta \leq \varepsilon \} {\rm ~from~} \beta=- q_{1} {\rm ~to~} \beta=0 {\rm ~along~ the~ imaginary~ line~ of~} z.
\end{array}
$$
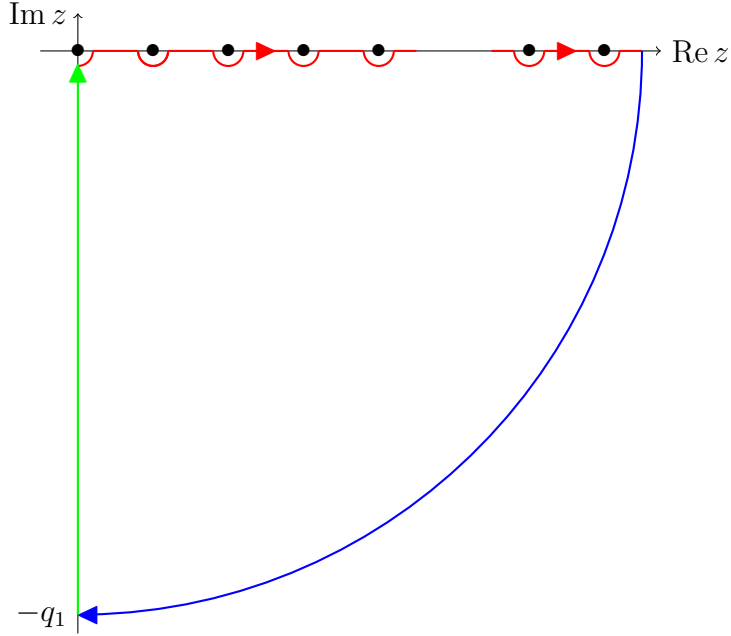
\begin{figure}
  \centering
\begin{tikzpicture}

\draw[->] (-0.5,0) -- (7.75,0) node[right] {${\rm Re\,}z$};
\draw[->] (0,-7.75) --(0,0.5) node[left] {${\rm Im\,}z$};
\draw[red,thick] (0,-0.2) arc (-90:0:0.2);
\draw[red,thick] (0.2,0)-- (0.8,0);
\draw[red,thick] (0.8,0) arc (-180:0:0.2);
\draw[red,thick] (1.2,0)-- (1.8,0);
\draw[red,thick] (1.8,0) arc (-180:0:0.2);
\draw[red,thick] (2.2,0)-- node {$\blacktriangleright$} (2.8,0);
\draw[red,thick] (2.8,0) arc (-180:0:0.2);
\draw[red,thick] (3.2,0)-- (3.8,0);
\draw[red,thick] (3.8,0) arc (-180:0:0.2);
\draw[red,thick] (4.2,0)-- (4.5,0);
\draw[red,thick] (0.8,0) arc (-180:0:0.2);
\foreach \Point in {(0,0), (1,0), (2,0), (3,0), (4,0)}{\node at \Point {\textbullet};};

\foreach \Point in {(6,0), (7,0)}{\node at \Point {\textbullet};};
\draw[red,thick] (5.5,0)-- (5.8,0);
\draw[red,thick] (5.8,0) arc (-180:0:0.2);
\draw[red,thick] (6.2,0)-- node {$\blacktriangleright$} (6.8,0);
\draw[red,thick] (6.8,0) arc (-180:0:0.2);
\draw[red,thick] (7.2,0)-- (7.5,0);
\draw[blue,thick] (7.5,0) arc (0:-89:7.5)node {$\blacktriangleleft$};

\node[left] at (0, -7.5) {$-q_1$};
\draw[->,green,thick] (0, -7.5) -- (0,-0.3) node {$\blacktriangle$};
\end{tikzpicture}
\caption{The contour ${\cal C}$ composed out of the three parts ${\cal C}_1$ (red), ${\cal C}_2$ (blue) and ${\cal C}_3$ (green).}\label{CFigure}
\end{figure}
By doing so, we interpret the $\sigma$-integral of (\ref{gosc1}) as
\begin{equation}\label{gosc2}
\mathcal{G}^{(Osc)}_\ell(r'', r'; E_{b}) = \lim_{\varepsilon \rightarrow 0, \, q_{1} \rightarrow \infty} \, \int_{{\cal C}_{1}}\rmd z  F(z).
\end{equation}
Since the contribution of the integral along ${\cal C}_{2}$ diminishes as $q_1$ tends to infinity, Cauchy's theorem implies that
\begin{equation}\label{gosc3}
\fl
\begin{array}{l}
\mathcal{G}^{(osc)}_{\ell}(\rho'', \rho'; E_{b}) = \displaystyle - \int_{{\cal C}_{3}}\rmd z F(z) \\[4mm]
\qquad= \displaystyle \frac{m\sqrt{\rho'\rho''}}{\hbar^{2}}
\int_{0}^{\infty}\frac{\rmd q}{\sinh q}
\exp\left\{- \frac{m\omega\left(\rho'^{2} + \rho''^{2} \right)}{2\hbar \tanh q}\right\}\rme^{(E_{b}/\hbar\omega)q}
\, I_{\ell + \frac{1}{2}} \left(\frac{m\omega \rho'\rho''}{\hbar \sinh q}\right)\,,
\end{array}\end{equation}
which is integrable.

At this point, we take $\mathcal{G}^{(Osc)}_{\ell}(\rho'', \rho'; E_{b})$ to $\mathcal{G}^{(Osc)}(\rho'', \rho'; L_{b}; E_{b})$ by changing $\ell+ 1/2$ back to $L_{b}$ in (\ref{gosc3}). Since the $\ell$th Green function,
$\mathcal{G}^{(Osc)}_{\ell}(\rho'', \rho'; E_{b})$, has no power-dual counterpart, we deal with the $L$-dependent Green function in order to proceed our argument on the base of power duality. As needed, we return to the $\ell$-dependent Green function by letting $L_{b}=\ell + 1/2$.

The result of integration on the right hand side of (\ref{gosc3}) can be found from formula 6.669.4 of the Gradshteyn-Ryzhik table \cite{Gradshteyn and Ryzik}.  After changing the integration variable from $x$ to $q$ by $\coth (x/2) = \rme^{q}$ it reduces to the formula,
\begin{equation}\label{GR66694}
\fl
\begin{array}{l}
\displaystyle
\int_{0}^{\infty}\, \frac{\rmd q}{\sinh q}\, \exp\left\{\frac{-\alpha  (a_1 + a_2)}{\tanh q}\right\} \rme^{2\nu q} \, I_{2\mu}\left(\frac{2\alpha\sqrt{a_1a_2}}{\sinh q}\right) \\[4mm]
\displaystyle
 \qquad\qquad\qquad  =  \frac{\Gamma(\mu - \nu + \frac{1}{2}) }{2\alpha \sqrt{a_1a_2}\, \Gamma(2 \mu + 1)} W_{\nu, \mu}(2\alpha a_1) M_{\nu, \mu}(2\alpha a_2)
\end{array}
\end{equation}
where
${\rm Re\,}(\mu - \nu + 1/2) > 0$, ${\rm Re\,} \mu > 0$, and $a_1 > a_2$.
With the help of this formula, we obtain the one-dimensional $L$-dependent Green function for the radial oscillator,
\begin{equation}\label{Gosc4}
\fl
\begin{array}{l}
\mathcal{G}^{(osc)}(\rho'', \rho'; L_{b};  E_{b})  \\
\qquad\qquad
\displaystyle
=  \frac{1}{\hbar \omega \, \sqrt{\rho'\rho''}} \, \frac{\Gamma\left(\frac{1}{2}L_{b} - k_{b} +\frac{1}{2}\right)}{\Gamma\left(L_{b} + 1\right)}
W_{k_{b}, \frac{1}{2}L_{b}}\left(\frac{m\omega}{\hbar}\rho''^{\,2}\right) M_{k_{b}, \frac{1}{2}L_{b}}\left(\frac{m\omega}{\hbar}\rho'^{\,2}\right)
\end{array}\end{equation}
where $k_{b}=E_{b}/(2\hbar\omega)$ and $\rho'' > \rho'$. By letting $L_{b}=\ell + \frac{1}{2}$ back, we have the
$\ell$-dependent Green function,
\begin{equation}\label{OGell}
\mathcal{G}^{(Osc)}_{\ell} (\rho'', \rho'; \, E_{b})= \mathcal{G}^{(osc)}(\rho'', \rho'; L_{b};  E_{b})|_{L_{b}=\ell + \frac{1}{2}}.
\end{equation}
Since $\varphi^{(b)}(\rho'\rho'')=\frac{1}{2}\sqrt{\rho'\rho''}$,  the corresponding dual-symmetric Green function is given by
\begin{equation}\label{Gosc5}
\fl
\begin{array}{l}
\mathbf{G}^{(osc)}(\rho'', \rho'; L_{b};  E_{b})=
\displaystyle\frac{1}{\varphi(\rho'\rho'')}\mathcal{G}^{(osc)}(\rho'', \rho'; L_{b};  E_{b})  \\[4mm]
\qquad\qquad\qquad=\displaystyle \frac{2}{\hbar \omega \, \rho'\rho''} \, \frac{\Gamma\left(\frac{1}{2}L_{b} - k_{b} +\frac{1}{2}\right)}{\Gamma\left(L_{b} + 1\right)}
W_{k_{b}, \frac{1}{2}L_{b}}\left(\frac{m\omega}{\hbar}\rho''^{\,2}\right) M_{k_{b}, \frac{1}{2}L_{b}}\left(\frac{m\omega}{\hbar}\rho'^{\,2}\right)\,.
\end{array}
\end{equation}
In this manner, we have derived the Green function from the path integral result (\ref{OsciK}) for the oscillator. Of course, the Green function (\ref{Gosc4}) provides us the energy spectrum from its poles and the eigenfunctions from the residues at the poles.

In general, the Gamma function can be separated to a singular part and a non-singular part as
\begin{equation}\label{Gamma1}
\Gamma(\zeta)=\sum_{n=0}^{\infty}\frac{(-1)^{n}}{n! \,(\zeta + n)} + \Phi(\zeta)
\end{equation}
where $\Phi(\zeta)$ is an integral function.
Evidently it has simple poles at $\zeta = -n$ with residues given by $\frac{(-1)^{n}}{n!}$. Hence $\Gamma\left(\frac{1}{2}L_{b} - k_{b} +\frac{1}{2}\right)$ in the numerator of (\ref{Gosc4}) may be represented by the dominant term near a pole $k_{b}=k_{b, n}$ for any
$n \in \mathbb{N}_{0}$ where $k_{b,n} =n + (L_{b} + 1)/2$,
\begin{equation}\label{Gamma2}
\Gamma\left(\frac{1}{2}L_{b} - k_{b} +\frac{1}{2}\right)\approx \frac{(-1)^{n}}{n!}\frac{1}{k_{b, n} - k_{b}}
= 2\hbar \omega \frac{(-1)^{n}}{n!}\frac{1}{E_{n}^{(Osc)} - E_{b}}
\end{equation}
where $E_{n}^{(Osc)}=2\hbar \omega \, k_{b, n}$. The pole at $E_{b}=E_{n}^{(Osc)}$ of (\ref{Gamma2}) is indeed the energy spectrum for the radial oscillator, if $L_{b}=\ell + \frac{1}{2}$,
\begin{equation}\label{Ospec}
E^{(Osc)}_{n, \ell}=\hbar \omega \left(2n + \ell + \frac{3}{2}\right)
\end{equation}
where $n\in \mathbb{N}_{0}$ and $\ell \in \mathbb{N}_{0}$.
Noticing also that, when $\nu = \frac{\alpha + 1}{2} + n$ and $\mu=\alpha/2$,  the Whittaker function $W_{\nu, \mu}(z)$ is expressible in terms of $M_{\nu, \mu}(z)$ as
\begin{equation}\label{WM}
W_{\frac{\alpha +1}{2} + n, \, \frac{\alpha}{2}}(z) = (-1)^{n} \frac{\Gamma(n + \alpha + 1)}{\Gamma(\alpha + 1)} M_{\frac{\alpha + 1}{2} + n,\, \frac{\alpha}{2}}(z),
\end{equation}
we have the residues of the Green function (\ref{Gosc4}),
\begin{equation}\label{Oresi}
\begin{array}{l}
\fl
\displaystyle
\mathrm{Res}\, _{E_{b}=E_{n}^{(osc)}} \, \mathcal{G}^{(osc)}(\rho'', \rho'; L_{b};  E_{b}) = v_{n}^{\ast}(\rho') \, v_{n}(\rho'')  \\
\displaystyle
  = \frac{2}{ \sqrt{\rho'\rho''}} \, \frac{\Gamma\left(n + L_{b} + 1\right)}{n!\, [\Gamma\left(L_{b} + 1\right)]^{2}}
 M_{k_{b, n}, \, \frac{1}{2}L_{b}}\left(\frac{m\omega}{\hbar}\rho''^{\,2}\right)
 M_{k_{b, n}, \, \frac{1}{2}L_{b}}\left(\frac{m\omega}{\hbar}\rho'^{\,2}\right)
\end{array}
\end{equation}
from which we can read off the normalised energy eigenfunctions defined in $L^{2}(\rmd\rho)$ space\footnote{The eigenfunctions (\ref{Owave}) are normalised by
$$
\int_{0}^{\infty}\rmd\rho \,|v_{n}(\rho)|^{2} = 1.
$$
This can easily be confirmed with the help of the formula,
$$
\int_{0}^{\infty} \rmd x\,  [M_{\frac{\alpha + 1}{2} + n, \frac{\alpha}{2}}(x)]^{2} \,x^{-1} =
\frac{n!\, [\Gamma(\alpha + 1) ]^{2}}{\Gamma(n + \alpha + 1)}\,,
$$
obtainable from the orthogonality relation of the Laguerre polynomials.}
by letting $L_{b}=\ell + 1/2$,
\begin{equation}\label{Owave}
v_{n, \ell}(\rho)= \sqrt{\frac{2\Gamma\left(n + \ell + 3/2\right)}{n!\, [\Gamma\left(\ell + 3/2\right)]^{2}}\,} \,
\rho^{-1/2}\,M_{n + \frac{1}{2}\ell + \frac{3}{4},\,  \frac{1}{2}\ell +\frac{1}{4}}\left(\frac{m\omega}{\hbar}\rho^{\,2}\right).
\end{equation}

So far we have examined the process for reproducing the solutions of the Schr\"odinger equation from the path integral result via the Green function for the oscillator. However, power duality has played no role. Next we turn ourselves to the power-dual partner, the Coulomb system. Since no closed form expression of the propagator is available for the hydrogen atom, we attempt
to construct the dual-symmetric Green function $\mathbf{G}^{(Coul)} (r'', r'; L_{a};  E_{a})$ out of $\mathbf{G}^{(Osc)}(\rho'', \rho'; L_{b};  E_{b})$ via the power-dual formula (\ref{GaGb}).  The duality transformation $\Delta_\delta$ with $\eta=2$ for $a=-1$ and $b=2$ consists of
$$
\begin{array}{l}
\mathfrak{R}: \,\, r=C\rho^{2}\,, \qquad \delta \mathfrak{T}: \,\, \delta t =4C^{2} \rho^{2} \delta s\,,  \qquad \,  \mathfrak{L}: \,\, L_{b} =2L_{a}\,,\\
\mathfrak{E}: \,\, E_{b} = - 4C \lambda_{a}\,, \quad \lambda_{b} = - 4C^{2} E_{a}\,, \qquad \mbox{and}
\qquad \mathfrak{S}: \, \, \lambda_{b'}=0\,.
\end{array}
$$
Before replacing the quantities with subscript $b$ by those with $a$ in the Green function (\ref{Gosc4}), we fix the constant $C$ in operation $\mathfrak{R}$ by using the second relation of operation $\mathfrak{E}$; namely $C=\sqrt{\lambda_{b}/(-4E_{a})\,}$ or
$C=m\omega / (2\kappa \hbar)$ where $\kappa\hbar = \sqrt{-2mE_{a}\,}$. Then $\mathfrak{R}$ implies $(m\omega/\hbar)\rho^{2}=2\kappa r$, and the first relation of $\mathfrak{E}$ yields $k_{b}=E_{b}/(2\hbar \omega)=mZe^{2}/(\kappa \hbar^{2})$.
Finally, letting $k_{a}=mZe^{2}/\hbar \sqrt{-2mE_{a}}$ and $L_{b}=2L_{a}=\ell + 1/2$ with $\ell \in \mathbb{N}_{0}$, we arrive at the $L$-dependent symmetrized radial Green function for the hydrogen-like atom,
\begin{equation}\label{Gcou1}
\begin{array}{l}
\fl\mathbf{G}^{(Coul)}_{\ell}(r'', r';L_a; E_{a};\lambda_a)=\mathbf{G}^{(Osc)} \left(\sqrt{r''/C}, \sqrt{r'/C}; 2L_{a}; - 4C \lambda_{a};-4C^2E_a\right)  \\
 \displaystyle = \frac{m}{\kappa\hbar^{2} \sqrt{r' r''}}  \, \frac{\Gamma\left(L_{a} - k_{a} + \frac{1}{2} \right)}{\Gamma\left(2L_{a} + 1 \right)}
W_{k_{a}, L_{a}}\left(2\kappa\, r''\right) M_{k_{a}, L_{a}}\left(2\kappa\, r' \right).
\end{array}
\end{equation}
With $\varphi(r'r'')=\sqrt{r'r''}$, the $L$-dependent one-dimensional radial Green function for the Coulomb system takes the form,
\begin{equation}\label{Gcou2}
\fl
\begin{array}{rl}
\mathcal{G}^{(Coul)}(r'', r'; L_{a}; E_{a})&=\varphi^{(a)}(r'r'')\,\mathbf{G}^{(Coul)}(r'', r';L_a;  E_{a}) \\
 &\displaystyle = \frac{m}{\kappa\hbar^{2}}  \, \frac{\Gamma\left(L_{a} - k_{a} + \frac{1}{2} \right)}{\Gamma\left(2L_{a} + 1\right)}
W_{k_{a}, L_{a}}\left(2\kappa\, r''\right) M_{k_{a},L_{a}}\left(2\kappa\, r' \right)\,,
\end{array}
\end{equation}
where $k_{a}=mZe^{2}/(\kappa \hbar^{2})$.  To convert the $L$-dependent Green function to the $\ell$-dependent Green function, we have let $L_{b}=\ell + \frac{1}{2}$ for the oscillator case. To move from the oscillator to the hydrogen atom, we have demanded $L_{b}=2L_{a}$. We have not demanded $\ell_{b} + \frac{1}{2} = 2\left(\ell_{a} + \frac{1}{2}\right)$ because this relation is wrong. The power-dual symmetry breaks down when the angular momentum is quantized. By the quasi-duality procedure, once the dual transformation is completed between the $L$-dependent quantities, we let $L_{a}=\ell + \frac{1}{2}$ as well as $L_{b}=\ell + \frac{1}{2}$.
With $L_{a}=\ell + \frac{1}{2}$, the $\ell$ dependent radial Green function for the hydrogen-like atom in $3$-dimensional space is found as
\begin{equation}\label{Gcou3}
\fl
\begin{array}{rl}
G^{(Coul)}_{\ell}(r'', r'; E_{a})&=(r'r'')^{-1} \mathcal{G}^{(Coul)}(r'', r';  L_{a}; E_{a}) |_{L{a}=\ell + \frac{1}{2}}  \\
& \displaystyle=  \frac{m}{\kappa\hbar^{2}r'r''}  \, \frac{\Gamma\left(\ell - k_{a} + 1 \right)}{\Gamma\left(2\ell + 2 \right)}
W_{k_{a}, \ell + \frac{1}{2}}\left(2\kappa\, r''\right) M_{k_{a}, \ell + \frac{1}{2}}\left(2\kappa\, r' \right).
\end{array}
\end{equation}

The Gamma function in the numerator of (\ref{Gcou2}) has poles at $k_{a}=k_{a, n}$ where
$k_{a, n}=n + L_{a} + \frac{1}{2}$ with $n \in \mathbb{N}_{0}$, and can be expressed near the pole at $k_{a}=k_{a, n}$ for any $n \in \mathbb{N}_{0}$ as
\begin{equation}\label{HGamma}
\Gamma (L_{a} - k_{a} + \frac{1}{2}) \approx \frac{ (-1)^{n}}{n!} \frac{1}{k_{a, n} - k_{a}} = \frac{(-1)^{n}}{n!\, k_{a, n}}
\frac{(\kappa_{n} \hbar)^{2}}{m} \frac{1}{E_{n}^{(Coul)} - E_{a}},
\end{equation}
where $\kappa_{n} \hbar = mZe^{2}/k_{a, n}$ and $E_{n}^{(Coul)} = - (\kappa_{n}\hbar)^{2}/(2m)$. Apparently, the poles of the Gamma function yield, if $L_{a}=\ell + \frac{1}{2}$, the energy spectrum of the hydrogen-like atom,
\begin{equation}\label{Hspec}
E^{(Cou)}_{n}= - \frac{ Z^{2}e^{4}m}{2\hbar ^{2} \tilde{n}^{2}}
\end{equation}
where $\tilde{n}=n +\ell + 1$ is the principal quantum number. At the pole $E_{a}=E_{n}^{(Coul)}$ for any $n \in \mathbb{N}_{0}$, the residue of the Green function (\ref{Gcou2}) reads
\begin{equation}\label{Hresi}
\begin{array}{l}
\fl
\mathrm{Res}\, _{E_{a}=E_{n}^{(Coul)}} \, \mathcal{G}^{(Coul)}(r'', r'; L_{a};  E_{a}) = u_{n}^{\ast}(r') \, u_{n}(r'')  \\
\displaystyle
 = \frac{\kappa}{k_{a, n}\, n!} \, \frac{\Gamma\left(n + 2L_{a} + 1\right)}{n!\, [\Gamma\left(2L_{a} + 1\right)]^{2}}
M_{k_{a, n}, \, L_{a}}\left(2\kappa r''\right)
M_{k_{a, n}, \, L_{a}}\left(2\kappa r'\right) ,
\end{array}
\end{equation}
which gives us the normalised energy eigenfunction defined in $L^{2}(\rmd r)$ space\footnote{The eigenfunctions (\ref{Hwave}) are normalised by
$$
\int_{0}^{\infty} \rmd r\,|u_{n}(r)|^{2} = 1.
$$
This normalisation is carried out with the help of the formula,
$$
\int_{0}^{\infty} \rmd x \, [M_{\frac{\alpha + 1}{2} + n, \frac{\alpha}{2}}(x)]^{2} =
\frac{n! \, [\Gamma(\alpha + 1) ]^{2}}{\Gamma(n + \alpha + 1)} \,(2n + \alpha + 1).
$$
}
with $L_{a}=\ell + 1/2$,
\begin{equation}\label{Hwave}
u^{(Cou)}_{n, \ell}(r) =\sqrt{\frac{\kappa \,\Gamma(n+ 2\ell + 2)\, }{n! \, (n + \ell + 1)\, [\Gamma(2\ell +2)]^{2}\, }} \, M_{n + \ell + 1, \, \ell + \frac{1}{2}}(2\kappa r).
\end{equation}

In this subsection, we have explicitly shown how we utilize the idea of power duality to construct  the Green functions and to reproduce the solutions of the Schr\"odinger equation for the hydrogen-like atom from the path integral results for the radial oscillator.
In the earlier paper [1], we have mentioned that the wave function of the hydrogen atom can be determined from that of the radial oscillator
\emph{except the normalisation factor}. It is remarkable that the path integral treatment with the help of power duality can produce the energy eigenfunctions \emph{including normalisation}.

\subsection{Example 2: \, A Family of Confinement Potentials $(a, a'; b, b')=(\frac{a'}{2}-1, a'; 0, 2)$}

As the second example, we take up a family of confinement potential models parameterized by the secondary parameter $a'$. The special case $a'=1$ was discussed in the previous paper \cite{IJ2021}. Let system $A$ be a particle bound in the zero-energy state $E_{a}=0$ in the confinement potential,
\begin{equation}\label{ConA}
V_{a}(r)=\lambda_{a}r^{\frac{a'}{2}-1} + \lambda_{a'}r^{a'}\qquad\mbox{with}\qquad \lambda_{a} < 0\, , \, \lambda_{a'} > 0\,.
\end{equation}
Obviously for $a' > 0$ this potential is of the confinement-type exhibiting bound states only. The particular choice $a'=1$ would indeed be a linear confinement potential for large $r$. In order to accommodate a zero energy eigenvalue we need to have for small $r$ an attractive potential. This leads us to the additional condition $a' \leq 2$. Hence, we will consider the range $0< a'\leq 2$ for our family of confinement potentials.

It is evident that for the special case $L_a=\frac{1}{2}$, which corresponds to $\ell = 0$, and the special choice $\lambda_a = - \hbar {a'}\sqrt{\lambda_{a'}/8m}$, the above potential generates a supersymmetric Witten model with SUSY potential $\Phi(r)=\sqrt{\lambda_{a'}}\,r^{a'/2}$ resulting in unbroken SUSY and hence a zero energy ground state \cite{Cooper1995,Junker2019}. For non-vanishing angular momentum $\ell > 0$ we even need a stronger attractive coupling constant. Hence, we require for the first coupling constant in (\ref{ConA}) the necessary condition
\begin{equation}\label{lambda_a}
  \lambda_a \leq - \sqrt{\frac{\hbar^2 {a'}^2\lambda_{a'}}{8m}}\,,
\end{equation}
which is not sufficient as will be seen below.

System $A$ has $a=\frac{a'}{2}-1$, so we have to apply $\mathfrak{R}$ with $\eta=2/(a+2)=4/(a'+2)$ and $b=-a\eta = 2(2-a')/(2+a'))$. Hence the duality transformation $\Delta_\delta$ of our choice consists of
\begin{equation}\label{powtrans2}
\mathfrak{R}: \, \quad \, r=f(\rho) =C\rho^{\frac{4}{2+a'}}
\end{equation}
\begin{equation}\label{dT2}
\delta \mathfrak{T}: \, \quad \, \delta t= \left(\frac{4}{2+a'}\right)^2C^{2} \rho^{\frac{4-2a'}{2+a'}} \delta s
\end{equation}
\begin{equation}\label{angtrans2}
\mathfrak{L}: \, \quad \, L_{b} = \frac{4}{2+a'} L_{a}.
\end{equation}
\begin{equation}\label{rename2}
\mathfrak{E}: \, \quad \, E_{b} = - \left(\frac{4}{2+a'}\right)^2 C^{(2+a')/2} \lambda_{a}, \, \quad \, \lambda_{b} = - \left(\frac{4}{2+a'}\right)^2C^{2} E_{a} \,.
\end{equation}
\begin{equation}\label{second2}
\mathfrak{S}: \, \quad  \lambda_{b'}= \left(\frac{4}{2+a'}\right)^2 C^{a'+2}\lambda_{a'}, \, \quad \, b'=\frac{2(a'-a)}{a+2} .
\end{equation}

Since $E_{a}=0$ is the choice we made for the confinement potential,  the second relation of $\mathfrak{E}$ demands $\lambda_{b}=0$, and the second relation of $\mathfrak{S}$ determines the power of the secondary potential $b'= 2$. The first relation of $\mathfrak{E}$ shows that system $B$ must have a positive energy $E_{b}$ as $\lambda_{a} < 0$. Therefore, system $B$ transformed from system $A$ by $\Delta_\delta$ has the modified potential
\begin{equation}\label{ConB}
U_{b}(\rho)=\lambda_{b'}r^{2}  - E_{b} \, \, \quad \, (\lambda_{b'} > 0, \, \, E_{b} > 0).
\end{equation}
Namely system $B$ turns out to be a radial harmonic oscillator with the energy $E_{b}$, for which the Green function has been evaluated in the previous example.  If we let $\lambda_{b'}=\frac{1}{2}m\omega^{2}$, the one-dimensional Green function for system $B$ is written as
\begin{equation}\label{GoscB}
\fl
\begin{array}{rl}
\mathcal{G}^{(osc)}(\rho'', \rho'; L_{b};  E_{b}) = & \displaystyle
- \frac{1}{\hbar\omega} \frac{\Gamma\left(\frac{1}{2}L_{b} - k_{b} +\frac{1}{2}\right)}{\Gamma\left(L_{b} + 1\right)}\\
\qquad\qquad\qquad& \displaystyle \times \frac{1}{\sqrt{\rho'\rho''}}\,
W_{k_{b}, \frac{1}{2}L_{b}}\left(\frac{m\omega}{\hbar}\rho''^{\,2}\right) M_{k_{b}, \frac{1}{2}L_{b}}\left(\frac{m\omega}{\hbar}\rho'^{\,2}\right),
\end{array}
\end{equation}
where $k_{b}=E_{b}/(2\hbar\omega)$. The poles of the Green function (\ref{GoscB}) occur when
$E_{b}=\hbar \omega (2\nu + L_{b} + 1)$, $\nu \in \mathbb{N}_{0}$.
With $\lambda_{b'}=\frac{1}{2}m\omega^{2}$, if we use the first relation of $\mathfrak{S}$ to fix the constant $C$ to be
\begin{equation}\label{OscC}
 C=\left(\frac{(2+a')^2 m\omega^{2}}{32 \lambda_{a'}}\right)^{\frac{1}{2+a'}},
\end{equation}
then the first relation of $\mathfrak{E}$ specifies a possible value out of the energy spectrum for system $B$ according to the chosen coupling constant $\lambda_{a}$ of the confinement potential,
\begin{equation}\label{ConEb}
E_{b} =\frac{\omega|\lambda_{a}|}{2+a'} \sqrt{\frac{8m}{\lambda_{a'}} }.
\end{equation}
Inserting the eigenvalues  $E_{b,\nu} = \hbar \omega (2\nu + L_{b} + 1)$ in above relation, results in admissible coupling constants given by
\begin{equation}\label{lambdaan}
  \lambda_{a,\nu} = -\left(\frac{a'+2}{4}\right)\sqrt{\frac{2\lambda_{a'}\hbar^2}{m}}\left(2\nu+\frac{4 L_a}{a'+2}+1\right)\,,
\end{equation}
which obviously obeys the upper bound (\ref{lambda_a}).

As the harmonic oscillator Green function (\ref{GoscB}) of system B is mapped via the duality transformation into that of system A at fixed energy $E_a=0$, so are the poles. In other words, for the admissible coupling constants (\ref{lambdaan}), the pole at $E_{b,\nu}$ is mapped onto the pole at $E_a=0$. Similarly, the residua are mapped onto each other. That is, the $\nu$-th eigenfunction of system B is mapped onto the zero-energy eigenfunction of system A and is given by
\begin{equation}\label{psin}
  \psi^{(a)}_{\nu\ell}(r) ={\cal N}_{\nu\ell} \,\, r^{-(1+\frac{a'}{4})} \,M_{\nu + \frac{2\ell +1}{2+a'}+ \frac{1}{2},\frac{2\ell +1}{2+a'}}\left(\alpha r^{\frac{2+a'}{2}}\right)
\end{equation}
with
\begin{equation}\label{Na'}
  \alpha = \frac{4}{2+a'}\frac{\sqrt{2m\lambda_{a'}}}{\hbar}
\end{equation}
and ${\cal N}_{\nu\ell}$ denotes the normalisation constant.
The corresponding effective potential reads
\begin{equation}\label{Veff}
  V^{eff}_{a,\nu}= \frac{(L_a^2-\frac{1}{4})\hbar^2}{2mr^2} - |\lambda_{a,\nu}| r^{a'/ 2- 1} + \lambda_{a'} r^{a'}.
\end{equation}
\begin{figure}[t]
\includegraphics[scale=0.75,bb=0 0 0 500]{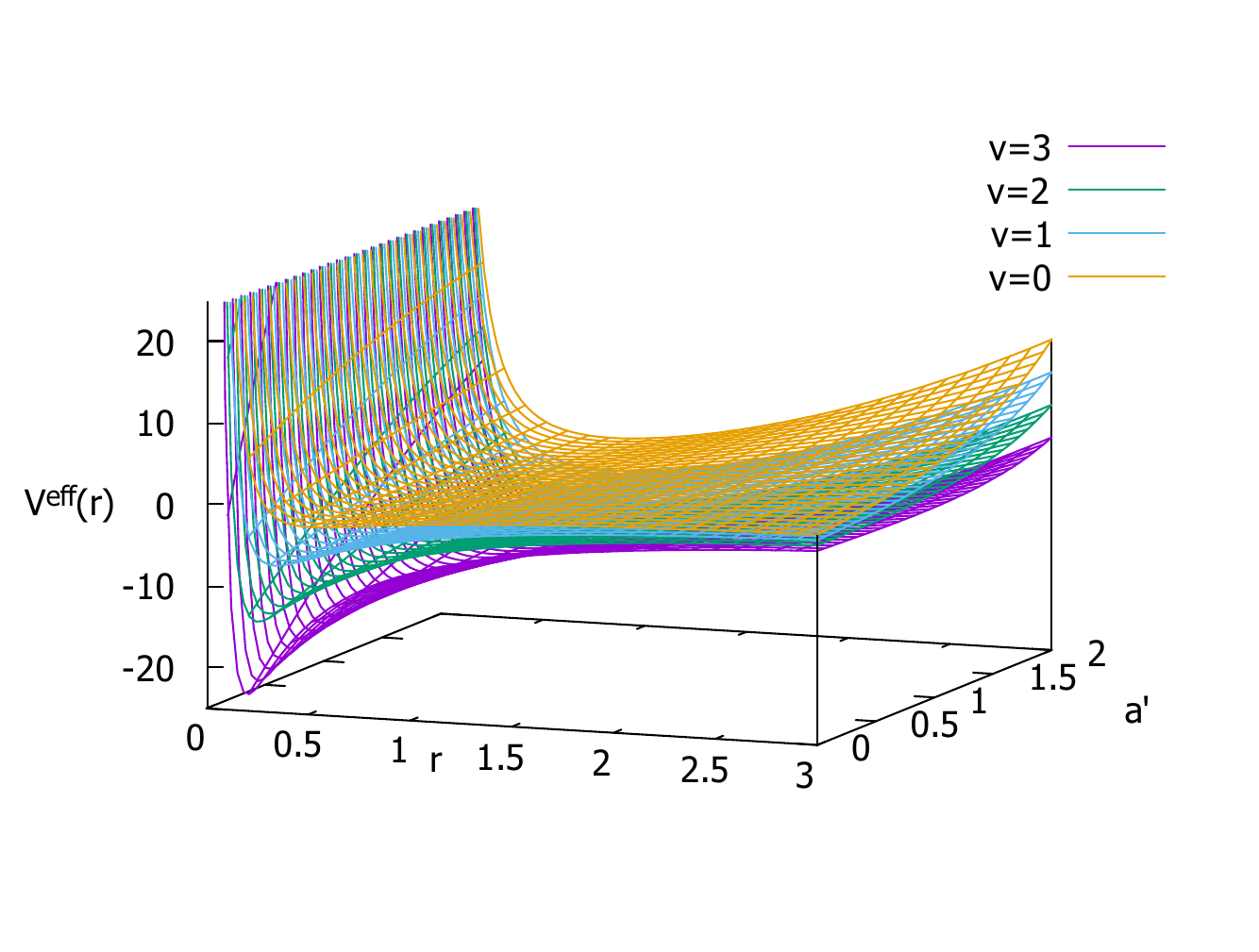}
\caption{The effective potential (\ref{Veff}) for $L_a=\frac{3}{2}$ in units where $\hbar=m= 1$ and $\lambda_{a'}=2$, for various values of the parameter $a'$ and for $\nu =0,1,2,3$ as indicated in the legend.}%
\end{figure}
In Figure 2 we show this effective potential for parameter $a'$ in the range $0<a'<2$ using units where $\hbar=m= 1$, $\lambda_{a'}=2$, angular momentum $L_a=3/2$ (this corresponds to $\ell_a=1$) and for values $\nu=0,1,2,3$. Obviously all these potentials generate a zero-energy eigenvalue for the associated Hamiltonian. This zero eigenvalue corresponds to the $\nu$-th energy eigenstate of system  A. Note that for increasing $\nu$ the parameter (\ref{lambdaan}) decreases and generates a lower minimum for the effective potential allowing also for negative eigenvalues.

In concluding this section let us look at some special case of the constructed family of potentials $\{V^{eff}_{a,n}\}$ which are of particular interest:

\noindent{\it Case $a'=0$}: \\
Despite the fact that we excluded that case in above discussion, it is obvious that $a'=0$ corresponds to the Coulomb problem discusses in the previous section. This case is exactly path integrable as discussed above. No limitation to $E_a\to 0$ is needed.

\noindent{\it Case $a'=2/3$}:\\
 In this case the effective potential reduces to a Stillinger-type potential of the form
\begin{equation}\label{Stillinger}
  V^{eff}_{a,n}(r)= \frac{(L_a^2-\frac{1}{4})\hbar^2}{2mr^2} - \sqrt{\frac{8\lambda_{a'}\hbar^2}{9m}}\left(2n+\frac{3}{2}L_a + 1\right) r^{-\frac{2}{3}} + \lambda_{a'} r^{\frac{2}{3}}.
\end{equation}
Here $a=-\frac{2}{3}$ and $\eta=\frac{3}{2}$. Let us note that this potential is similar in form to the one discussed by Stillinger \cite{Still1979}. Stillinger has an attractive centrifugal part with a specific coupling constant, which is not the case here. See also Ishkhanyan and Krainov \cite{Ishkhanyan2023}, where the Dirac problem for the $r^{1/3}$-potential is reduced to the non-relativistic Stillinger potential problem.

\noindent{\it Case $a'=1$}: \\
Here we have $a=-1/2$ and $\eta = 4/3$ and the potential is a linear confining one for large $r$,
which was extensively discussed in ref.\ \cite{IJ2021}.

\noindent{\it Case $a'=2$}:\\
 Obviously for this case we have $a=0$ and $\eta =1$. That is, this is a trivial transformation and the A system is again the radial harmonic oscillator, which is exactly path integrable.

Let us  recall Figure 2, which clearly shows the transition from the Coulomb case to the harmonic oscillator case when $a'$ changes from value $0$ to $2$.
\section{Summary}
In the present work we have investigated the properties of the path integral for the promotor under the power duality transformation $\Delta$ defined in (\ref{powtrans})-(\ref{second}).

$\bullet$  Feynman's path integral is re-formulated by using the action of Hamilton's characteristic form in the place of the action in Hamilton's principal form. The resulting path integral is for the promotor rather than the propagator. The radial promotor expressed as a path integral can be made invariant under $\Delta$ provided the angular momentum is not quantized.

$\bullet$  The power duality is found to be primarily a classical notion, which breaks down at the level of angular quantization. The idea of quasi-duality is proposed by modifying the angular momentum operation $L_{b}=|\eta| L_{a}$ in an ad hoc manner as $ \ell_{a}=\ell_{b}=\ell \in \mathbb{N}_{0}$ rather than $\ell_{b} + (D-2)/2 = |\eta|\{\ell_{a} + (D-2)/2 \}$.

$\bullet$  The quasi-dual radial Green functions are constructed out of the dual pair of promotors. A formula is proposed to find the Green function for one of a dual pair by knowing the Green function for the other.

$\bullet$  As the first example, the Coulomb-Hooke duality is presented in details. The way going from the propagator for the radial harmonic oscillator to the Green function for the radial oscillator and the Green function for the hydrogen-like atom is explicitly shown. Although it is well-known that the energy spectrums and the eigenfunctions can be obtained from the Green functions, the proposed duality method enables us to determine even the normalization of the wave functions for the dual pair.

$\bullet$ As the second example, a family of confinement models is discussed in which a particle is bound at the zero-energy state in a two-term power potential whose powers are adjustable by a single real parameter $a' \in [0, 2]$. Choosing system $B$ to be the radial oscillator, we have determined the effective potential for the bound system $A$, which interpolates between the Coulomb system ($a'=0$) and the Hooke system ($a'=2$) and includes the Stillinger-type potential ($a'=2/3$) as a special case.
The ground state eigenfunctions for the family of confined systems $A$ are obtained, and the condition on the admissible principal coupling constant is also found for the confining potential.

\appendix
\section{Details on the Angular Path Integration}
This appendix contains details of the angular path integration performed in section \ref{angularsubsection}. There
we introduce the $D$ dimensional unit vector, $\mathbf{u}=\mathbf{r}/r \in \mathbb{R}^{D}$ on the unit hypersphere $\mathbb{S}^{D-1}\subset \mathbb{R}^{D}$. Let its $k$-th component be parameterized by angular variables as
\begin{equation}\label{Apolcoords}
u^{(k)} = \prod_{i=0}^{k-1} \sin \theta^{(i)} \cos \theta^{(k)}\,,\qquad k=1,2,3,\ldots,D\,,
\end{equation}
where $\theta^{(0)}=\pi /2$, $\theta^{(D)}=0$, $0 \leq \theta^{(D-1)} \leq 2\pi $, and $0 \leq \theta^{(k)} \leq \pi$ for $k=1, 2,..., D-2$. Certainly $\mathbf{u}\cdot \mathbf{u}=\sum_{k=1}^{D} u^{(k)} u^{(k)} =1$.  The element of solid angle subtended by the surface element of the hypersphere $\mathbb{S}^{D-1}$ is given by
\begin{equation}\label{Adomega}
\rmd^{D-1}\Omega(\mathbf{u}) =\prod_{k=1}^{D-1}\rmd\theta^{(k)}\left(\sin \theta^{(k)}\right)^{D-1-k}
\end{equation}
which integrates to the total surface area of the hypersphere,
\begin{equation}\label{Avolume}
\int_{\mathbb{S}^{D-1}} \rmd^{D-1}\Omega(\mathbf{u})= \frac{2 \pi^{D/2}}{\Gamma(D/2)}.
\end{equation}

Now we define $\mathbf{u}_{j}$ for each $\mathbf{r}_{j}$ and write the integration measure in (\ref{Psliced1}) as
\begin{equation}\label{Ameasure}
\rmd^{D}\mathbf{r}_j=r_{j}^{D-1}\,\rmd r_{j}\,\,  \rmd^{D-1}\Omega(\mathbf{u}_{j}).
\end{equation}
Since $(\Delta \mathbf{r}_{j})^{2}=r_{j}^{2} + r_{j-1}^{2} - 2r_{j}r_{j-1} \mathbf{u}_{j}\cdot \mathbf{u}_{j-1}$ in polar variables, we can express the short time promotor (\ref{shortPro1}) as
\begin{equation}\label{AshortPro2}
\fl
\begin{array}{rl}
P(\mathbf{r}_{j}, \mathbf{r}_{j-1}; \tau_{j})=&\displaystyle\left[\frac{m}{2\pi \rmi \hbar \tau_{j}}\right]^{\frac{D}{2}}
\exp \left\{ \frac{\rmi m}{2\hbar \tau_{j}} \left(r_{j}^{2} + r_{j-1}^{2} \right) - \frac{\rmi}{\hbar}U_{j}\tau_{j} \right\}\\[2mm]
&\times\displaystyle \exp \left\{\frac{m \hat{r}^{2}_{j}}{\rmi \hbar \tau_{j}} \mathbf{u}_{j} \cdot \mathbf{u}_{j-1} \right\},
\end{array}\end{equation}
where $U_{j}=U(r_{j})$ and $\hat{r}^{2}_{j} = r_{j} r_{j-1}$. Here  $U(r)$ is the modified potential $U(r)=V(r) - E$.

To separate the radial variable and the angular variables mixed in the last exponential function of (\ref{AshortPro2}) we employ the Gegenbauer expansion formula 8.534 of table \cite{Gradshteyn and Ryzik},
\begin{equation}\label{AGform1}
\fl
\rme^{z \cos \vartheta} = \left(\frac{2}{z}\right)^{\mu} \Gamma(\mu) \sum_{\ell =0}^{\infty}
(\ell + \mu ) C_{\ell}^{\mu}(\cos \vartheta )\, I_{\ell+ \mu}(z)\,, \qquad \mu \neq 0, -1, -2, \dots \,,
\end{equation}
where $C_{\ell}^{\mu}(x)$ is the Gegenbauer polynomial and $I_{\ell + \mu}(x)$ is the modified Bessel function of first kind. For $\cos \vartheta = \mathbf{u}_{j}\cdot \mathbf{u}_{j-1}$, and $\mu = (D-2)/2$, we can put (\ref{AGform1}) into the form,
\begin{equation}\label{AGform2}
\rme^{z \mathbf{u}_{j}\cdot \mathbf{u}_{j-1}} = 2\pi \left(\frac{2 \pi}{z}\right)^{(D-2)/2} \sum_{\ell_{j} =0}^{\infty}
\mathcal{C}_{\ell_{j}}^{(D-2)/2}(\mathbf{u}_{j}\cdot \mathbf{u}_{j-1})\, I_{\ell_{j}+ (D-2)/2}(z)
\end{equation}
where $\mathcal{C}_{\ell}^{(D-2)/2}(\mathbf{u}\cdot \mathbf{u}')$ is the modified Gegenbauer polynomial defined by
\begin{equation}\label{AGmodif}
\mathcal{C}^{(D-2)/2}_{\ell}(\mathbf{u}' \cdot \mathbf{u})=\frac{(2\ell + D - 2)\Gamma(D/2)}{2(D-2) \pi^{D/2}}\,C^{(D-2)/2}_{\ell}(\mathbf{u}'\cdot \mathbf{u})
\end{equation}
which obey the orthonormality relation (\ref{GGG}). Note that the modified Gegenbauer polynomials can be given in terms of the hyperspherical harmonics,
$\mathcal{Y}_{\ell}^{m}(\mathbf{u})$, satisfying the orthonormality relations,
\[
\int \, \rmd^{D-1}\Omega (\mathbf{u})\, \, \mathcal{Y}_{\ell }^{\rm m^{ \ast}}(\mathbf{u}) \, \mathcal{Y}_{\ell'}^{\rm m'}(\mathbf{u}) = \delta_{\ell, \ell'} \delta_{m, m'}.
\]
Namely,
\[
\mathcal{C}_{\ell}^{(D-2)/2}(\mathbf{u}'' \cdot \mathbf{u}')= \sum_{\rm m =1}^{M} \mathcal{Y}_{\ell}^{\rm m^{\ast}}(\mathbf{u}'') \, \,
\mathcal{Y}_{\ell}^{\rm m }(\mathbf{u}')
\]
where $M=(2\ell + D - 2) (\ell + D - 3)!/[\ell! (D-2)!]$. See, e.g., ref.\ \cite{Avery2010}.
Inserting $z=m\hat{r}^{2}_{j}/(i\hbar \tau_{j})$ into (\ref{AGform2}) and using the series expansion (\ref{AGform2}) in (\ref{AshortPro2}), we obtain (\ref{shortPro3}).

\section{On the Approximation of the Radial Short Time Promotor}
For the purpose of comparison between the short time action and the classical action, we wish to bring the contribution from the modified Bessel function in the radial path integral (\ref{shortR1}) into the exponential form by utilizing the asymptotic behavior of the modified Bessel function for large $|z|$,
\begin{equation}\label{Basym}
I_{\mu}(z) \sim \frac{1}{\sqrt{2\pi z}} \left\{ \rme^{z} \sum_{k=0}^{\infty} (-1)^{k} \frac{(\mu, k)}{(2z)^{k}} + \rme^{-z} \rme^{\pm \rmi\pi (\mu + 1/2)}
\sum_{k=0}^{\infty} \frac{(\mu, k)}{(2z)^{k}} \right\}\,,
\end{equation}
where $(\mu, k)=\Gamma(\mu + k + 1/2) /[ k! \Gamma(\mu - k + 1/2)]$, and the $+$ sign in the second term is for $-\pi < {\rm Arg\,} z < 3\pi/2$ and the $-$ sign is for $-3\pi/2 < {\rm Arg\,} z < \pi/2$.  The first term dominates if ${\rm Re \,}z > 0$.

For the convergence of the Feynman measure, a couple of simple tricks have been proposed. Feynman's choice \cite{Feynman1948}
is that the Planck constant has a small negative part; Cameron \cite{Cameron} assumed that the mass $m$ has a small positive part. Another common choice is to apply Wick rotation from a real time to an imaginary time, i.e., $t \rightarrow -\rmi t$.
 See also a nice summary on this matter by Klauder \cite{Klauder2003}.
Under any one of the above choices,  ${\rm Re \,}z > 0$ for $z=m\hat{r}_{j}^{2}/(\rmi\hbar \tau_{j})$. Since $|z|$ is large (because $\tau_{j}$ is small), the modified Bessel function $I_{L}(z)$ with $L = \ell+(D-2)/2$, contained in the short time radial promotor (\ref{shortR1}), may be approximated by
\begin{equation}\label{BBessel}
I_{L}\left(\frac{m\hat{r}_{j}^{2}}{\rmi\hbar \tau_{j}} \right) \dot{=} \left(\frac{\rmi\hbar \tau_{j}}{2\pi m \hat{r}_{j}^{2}}\right)^{1/2}\, \exp\left\{\frac{m}{\rmi\hbar \tau_{j}} \hat{r}_{j}^{2} - \frac{\rmi}{\hbar} \frac{\left(L^{2} - \frac{1}{4}\right) \hbar^{2}}{2m \hat{r}_{j}^{2}}\tau_{j}\right\}.
\end{equation}
The symbol $\doteq $ used here signifies the approximate equality valid in a time-sliced path integral \cite{IKG1992}. This then directly leads us to the radial short time promotor (\ref{radprom2}).

\end{document}